\documentclass[10pt,twocolumn,twoside]{IEEEtran}
\usepackage{color}
\usepackage{graphicx,epsfig}
\usepackage{amsmath}
\usepackage{amssymb}
\usepackage{subfig}
\usepackage{tensor} 
\newcommand{\vertfig}[2][]{%
  \begin{minipage}{7in}\subfloat[#1]{#2}\end{minipage}}

\newtheorem{theorem}{Theorem}

\newtheorem{lemma}[theorem]{Lemma}

\newcommand\Perms[2]{\tensor[_{#1}]{\text{\it{P}}}{_{#2}}}

\normalsize

\ifCLASSINFOpdf
\else
\fi

\begin{document}
%
\title{Performance Analysis of Massive MIMO for Cell-Boundary Users}
%
%
%

\author{Yeon-Geun~Lim,~\IEEEmembership{Student~Member,~IEEE,}
        Chan-Byoung~Chae,~\IEEEmembership{Senior~Member,~IEEE,}
        and Giuseppe~Caire,~\IEEEmembership{Fellow,~IEEE}
\thanks{Y.-G. Lim and C.-B. Chae are with the School of Integrated Technology, Yonsei University, Korea. Email: \{yglim, cbchae\}@yonsei.ac.kr.}
\thanks{G. Caire is with the School of Electrical Engineering and Computer Science, Technical University of Berlin, Germany. Email: caire@tu-berlin.de.}}

%
%

\markboth{IEEE Transactions on Wireless Communications}%
{Minor Revision}
%



\maketitle

\begin{abstract}
In this paper, we consider massive multiple-input multiple-output (MIMO) systems for both downlink and uplink scenarios, where three radio units (RUs) connected via one digital unit (DU) support multiple {\color{black}user equipments (UEs)} at the cell-boundary through the same radio resource, i.e., the same time-frequency slot. {\color{black}For downlink transmitter options, the study considers zero-forcing (ZF) and maximum ratio transmission (MRT), while for uplink receiver options it considers ZF and maximum ratio combining (MRC). For the sum rate of each of these, we derive simple closed-form formulas. In the simple but practically relevant case where uniform power is allocated to all downlink data streams, we observe that, for the downlink, vector normalization is better for ZF while matrix normalization is better for MRT.} For a given antenna and user configuration, we also derive analytically the signal-to-noise-ratio (SNR) level below which MRC should be used instead of ZF. Numerical simulations confirm our analytical results. 
\end{abstract}


%
\IEEEpeerreviewmaketitle

\section{Introduction}\label{Section1}
%
%
%
%
Multiple-input multiple-output (MIMO) wireless communication techniques have evolved from single-user MIMO (SU-MIMO) to multi-user MIMO (MU-MIMO) systems \cite{Chae_SPMag_07}. To approach the capacity of the MIMO broadcast channel, \cite{SpencerSwindle04, Chae_JSAC07} proposed simple zero-forcing {\color{black}(ZF)-based} linear algorithms, where the transmitter and the receivers are equipped with multiple antennas. {\color{black}The authors in \cite{Chae_SPL09} intensively investigated the optimality of the linear matched type-combining filter and they assumed an infinite number of antennas at the receiver.} \cite{Chae_SPL09} proved that a simple linear beamforming (coordinated beamforming in the paper) asymptotically approaches the sum capacity achieved by dirty paper coding (DPC). 

Recently, massive MIMO (a.k.a. large-scale MIMO) has been proposed to further maximize network capacity and to conserve 
energy~\cite{Ngo_TCOM12, Marzetta_WC10, Rusek_SPMag_12}. {\color{black}The authors in \cite{Marzetta_WC10} proposed massive MIMO systems that use} simple linear algorithms such as {\color{black}maximum ratio combining (MRC)} for the uplink and {\color{black}maximum ratio transmission (MRT)} for the downlink. 
To further increase sum rate performances, several network MIMO algorithms with multiple receive antennas have been proposed~\cite{Chae_NCBF08, Chae_IACBF09}. These systems, {\color{black}however,} assume that the network supports a maximum of three users through a relatively small number of transmit antennas.\footnote{Note that more than three users can be supported if there is a common message, i.e., for a clustered broadcast channel.} 

Massive MIMO systems in multi-cell environments {\color{black}have also been} studied in~\cite{Ngo_TCOM12,Rusek_SPMag_12,Hoon11,Jubin_WC11}. 
Multi-cell massive MIMO {\color{black}is prone to} some critical issues. {\color{black}These include} pilot contamination, {\color{black}which becomes, in time division duplex (TDD) systems,} the main capacity-limiting factor, especially when MRT is used. Joint spatial division and multiplexing is proposed in \cite{Nam} to employ frequency division duplex (FDD) systems. The authors in \cite{Filippou} proposed a pilot alignment algorithm for a cognitive massive MIMO system. The authors in \cite{Rusek_SPMag_12} investigated downlink performance with MRT and ZF precoder for a massive MIMO system.\footnote{In \cite{Yang}, the authors investigated the performance of MRT and ZF in large-scale antenna systems, but {\color{black}paid little} attention to normalization techniques.} In~\cite{Ngo_TCOM12}, the authors studied uplink performance with MRC, ZF, {\color{black}and a} minimum mean square error (MMSE) {\color{black}filters} for massive MIMO. It was shown that transmit energy can be conserved by {\color{black}the} power-scaling law $1/M$ with perfect channel state information (CSI) and $1/\sqrt{M}$ with imperfect CSI at the base station (BS), where $M$ represents the number of BS antennas. In~\cite{Jubin_WC11}, the authors showed theoretically and numerically the impact of pilot contamination and proposed a multi-cell MMSE-based precoding algorithm to reduce both intra- and inter-cell interference. In~\cite{Jubin_WC11}, MRT precoding was used; the inter-user interference is eventually eliminated once the transmitter has a large enough number of antennas. 

The assumption of an infinite number of antennas at the BS for a finite number of users somehow trivializes many problems (e.g., {\color{black}under this limit} MRC/MRT have the same performance {\color{black}as} ZF). A more meaningful system scaling is considered in \cite{Hoon11,Antonia,Jakob}, where the number of antennas per BS and the number of users both go to infinity with {\color{black}a} fixed ratio. In this case, {\color{black}the results of} the infinite number of BS antenna{\color{black}s} per user can be recovered by letting this ratio become large. This more refined analysis, however, illuminates all the system performance regimes. For example, in \cite{Jakob} the ``massive MIMO'' regime is defined as the regime where pilot contamination dominates with respect to {\color{black}multi-user} interference{\color{black}; it is observed} that this regime occurs only when the number of BS antennas per user is impractically large.
These conclusions are also reached, independently and in parallel, in \cite{Hoon11}. In particular, \cite{Hoon11} considers a multi-cell architecture formed by small clusters of cooperating BSs. {\color{black}The authors} proposed a system where the users are partitioned into homogeneous classes and the downlink MIMO precoding scheme 
is optimized for each class. Then, a scheduler optimally allocates the time-frequency transmission resource across the different user classes, yielding an 
inherent multi-mode multi-cell massive MIMO system. One of the main {\color{black}findings} of \cite{Hoon11} is that it is convenient to serve users at the cell center in a 
single-cell mode, while users at the cell edge should be served by small cooperative clusters formed by the closest three neighboring cell/sectors.

{\color{black}This paper, motivated by the results in \cite{Hoon11}, focuses on the cell-edge user performance-the system bottleneck for both the uplink and the downlink.} 
The massive MIMO system under consideration consists of multiple radio units (RUs) connected {\color{black}to} one another by optical fibers, and further connected to a centralized digital unit (DU), as illustrated in Figs.~\ref{Fig:SysModel1}(a) and~\ref{Fig:SysModel2}.
Through the optical fibers, each RU can share data messages and channel state information. {\color{black}With respect to all three neighboring BSs forming a cluster, the {\color{black}cell-edge} users} have symmetric and spatially isotropic channel statistics. {\color{black}Hence,} the system is equivalent to, {\color{black}as shown in Fig.~\ref{Fig:SysModel1}(b)}, a single-cell massive MIMO system {\color{black}with {\color{black}cell-edge} users {\color{black}that are located in the low signal-to-noise-ratio (SNR) regime} {\color{black}(for the high SNR regime, we provide some results to provide a detailed analysis, such as for determining which normalization method is better for MRT)}}. 
As in the {\color{black}prior} work {\color{black}referenced above}, we consider the performance of linear precoding/filtering schemes such as ZF and MRT for the downlink and 
ZF and MRC for the uplink. In particular, we consider two possible normalization{\color{black}s} of the precoding filters for the downlink, referred to as vector {\color{black}and, respectively} matrix normalization. 
The main contributions of this paper are as follows:
\begin{quote}
$\bullet~$\emph{Tighter ergodic achievable sum rate of ZF and MRT/MRC}: {\color{black}Tighter ergodic achievable sum rate of ZF and MRT/MRC: We provide a methodology of simple but quite accurate approximations while considering the property of \emph{near deterministc}, which is defined in Section~\ref{Cheby}. This approximation method is valid for massive MIMO systems at low/high SNR. The effective channel matrix tends to an identity matrix when MRT/MRC are assumed with perfect CSI in massive MIMO systems. This means that the random channels become \emph{near deterministic} due to the property of the law of large numbers. If the number of user is large, the sum of the interference powers cannot become \emph{near deterministic}, which means the sum of those still have randomness and do not converge to zero. The researchers in \cite{Ngo_TCOM12} concluded that with perfect CSI at the BS and a large $M$, the performance of a MU-MIMO system with a transmit power per user scaling with $M$ is equal to the performance of a single-input-single-output system. Assuming a large number of users, the conclusion in \cite{Ngo_TCOM12} does not hold because the sum of the interference powers does not converge to zero.
From this aspect, considering the condition of \emph{near deterministic} and the large number
of users is mathematically significant in analyzing the performance of massive MIMO
systems. We investigate whether the signal-to-interference-plus-noise-ratio (SINR) term
in the ergodic sum rate is able to approximate \emph{near deterministic} such as the form
of the expectation when $M$ goes to infinity in the low or high SNR regime. From
this approximation of the SINR term, we derive simple approximations for the ergodic
achievable sum rate of ZF and MRT/MRC at the low and high SNR regimes considering
two simple power normalization methods at the downlink.}

The derived approximations are accurate and far simpler to evaluate than the {\em asymptotic expressions} given in \cite{Hoon11,Antonia,Jakob}, which were obtained through asymptotic random matrix theory \cite{RandomMatrixBook} and usually given in terms of the solution of multiple coupled fixed-point equations. 
Thanks to their simplicity, the proposed approximations are suitable to analyze the low and high SNR regimes. From the derived approximations, we investigate a suitable normalization method of MRT and transceiver mode selection algorithms. In their asymptotic expressions, the authors in \cite{Rusek_SPMag_12, Yang, Jakob} did not consider normalization methods; thus the expressions were unable to classify which normalization method was better.
Numerical results demonstrate the tightness of our analysis.\footnote{In fact, the achievable sum rate can be slightly enhanced in a low SNR regime by using regularized ZF (a.k.a. MMSE). This strategy, however, requires knowing all users noise variances at the transmitter, which requires additional feedback. Thus, in this paper we focus on ZF and/or MRT/MRC.}
\end{quote}
\begin{quote}
$\bullet~$\emph{Downlink precoding normalization methods}: We compare matrix and vector normalization for downlink precoding, in the simple and practical case of uniform power allocation over all downlink streams. It is {\color{black}well known} that {\color{black}these normalizations,} with optimal (waterfilling) power allocation and ZF precoding yield identical results \cite{Caire2010}. {\color{black}However}, with practical suboptimal power allocation and in the finite antenna regime these normalizations are generally not equivalent.
Most prior work on {\color{black}multi-user} MIMO paid {\color{black}scant} attention to this issue. {\color{black}For example, {\color{black}the authors in \cite{PerturPart1}, considered matrix normalization and those in \cite{PerturPart2} considered vector normalization in \cite{PerturPart2} with neither making} any mention of why different normalizations were used{\color{black};} \cite{PerturPart1} is part I of their papers dealing with linear precoding (ZF/MMSE), and \cite{PerturPart2} is part II focusing on non-linear precoding (vector perturbation). Thus, the comparison between vector and matrix normalization {\color{black}remains} an open problem. To solve this issue, we find a suitable normalization method for MRT precoding by using the proposed approximations and a suitable normalization method for ZF precoding by using the arithmetic-geometric inequality}. 
\end{quote}
\begin{quote}
$\bullet~$\emph{Transceiver mode selection algorithms}: We propose two transceiver mode selection algorithms {\color{black}from transmit power and the number of active users perspectives.}
In \cite{Louie}, it is concluded that ZF is better for cell center users, i.e., high SNR, and MRT is better for cell-boundary users, i.e., low SNR, in a downlink system.
However, \cite{Louie} did not consider transceiver mode selection as {\color{black}a} function of SNR (i.e., of the transmit power, for a given pathloss law and cell geometry) for the same class of edge users. In this paper, we explain how much transmit power and/or number of active users are needed for ZF to provide a better sum rate than MRT (for downlink) or MRC (for uplink). 
In particular, we find the optimal MIMO mode selection scheme in terms of closed-form thresholds of the transmit power, where the thresholds depend on the number of edge users. 
\end{quote}
Note that {\color{black}since our system model is simplified by making assumptions, our analytical results are more accurate than the work given demonstrated in \cite{Hoon11,Antonia,Jakob}}. {\color{black}Furthermore, {\color{black}our-closed} form expressions are based on {\color{black}the} assumption of infinite $M$, and {\color{black}when $M$ goes to infinity the approximation is quite accurate}. {\color{black}Even when $M$ is finite, though still a large number, the approximations are quite accurate.} In {\color{black}prior} work on massive MIMO with linear receivers/precoders, {\color{black}reserchers have} provided the simple lower bounds of the sum rate performance \cite{Ngo_TCOM12,Rusek_SPMag_12,Yang}, or the complex {\color{black}closed-form} expressions of that \cite{Jakob}, which are less accurate than our analysis.}
We {\color{black}anticipate} our contributions {\color{black}to yield} a wide range of insights {\color{black}for related studies, such as {\color{black}those on performance analysis on MIMO, power normalization methods, and the trade-off between ZF and MRT/MRC.}}

This paper is organized as follows. In Section~\ref{Section2}, we introduce the considered system model and problem statement with respect to precoding normalization 
methods and beamforming techniques. In Section~\ref{SectionPre}, we introduce some mathematical {\color{black}motivations} and preliminaries useful for analysis. 
In Section~\ref{Section3}, we analyze {\color{black}i) the ergodic performance of {\color{black}ZF and MRT precoding}, ii) which precoding normalization method is better for each precoder, and iii) {\color{black}the} ergodic performance for cell-boundary users with the best normalization method. In Section~\ref{Section4}, we provide an approximation of the achievable ergodic uplink sum rate.} In Section~\ref{Section5}, we propose transceiver mode selection algorithms with {\color{black}i) a power threshold and ii) the number-of-users} cross point of ZF- and {\color{black}MRT-precoding} techniques. Numerical results are shown in Section~\ref{Section6}. Section~\ref{Section7} presents our conclusions and future work.

\section{System Model: Massive MIMO}\label{Section2}

\begin{figure}[t]
 \centerline{\resizebox{1\columnwidth}{!}{\includegraphics{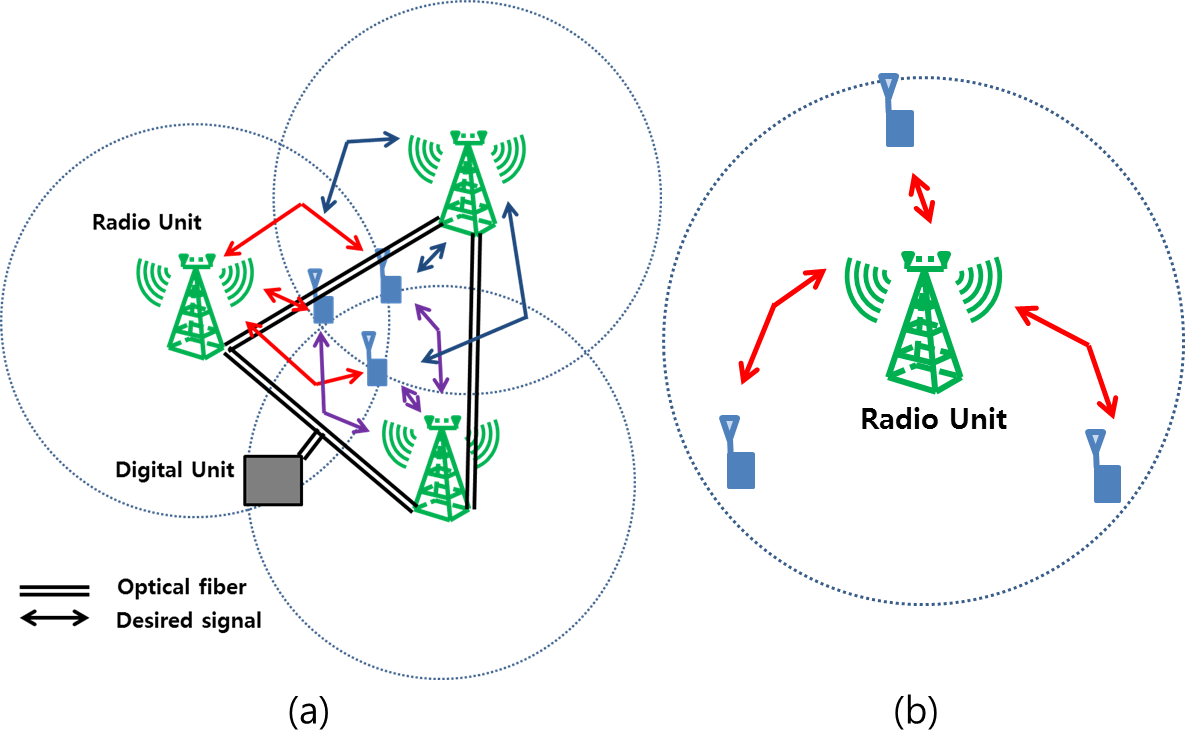}}}
  \caption{(a) System model of multi-RU massive MIMO scenario with cell-boundary users. (b) System model of single-RU massive MIMO scenario with cell-boundary users.}
  \label{Fig:SysModel1}
\end{figure}
\begin{figure}[t]
 \centerline{\resizebox{1\columnwidth}{!}{\includegraphics{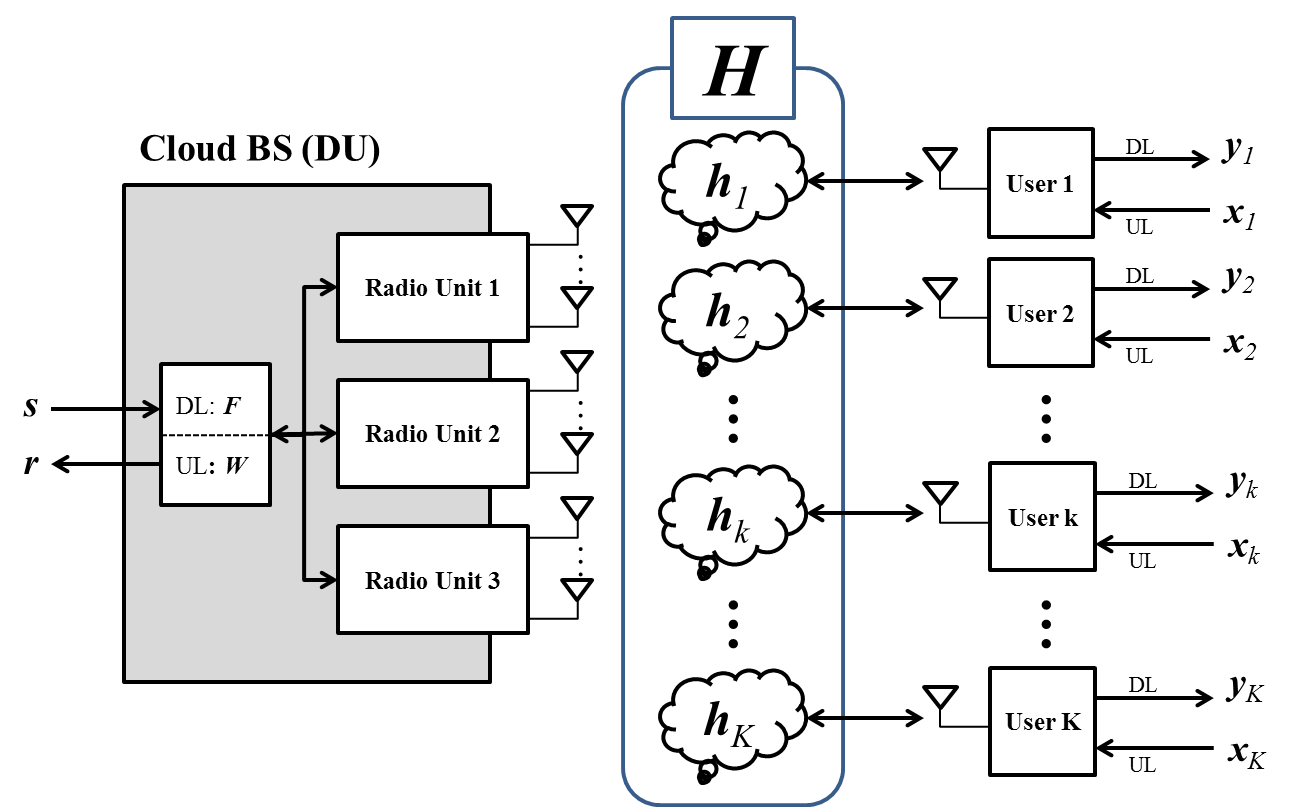}}}
  \caption{Block diagram of multi-RU massive MIMO system. One DU consists of three RUs connected by optical fibers. This model is regarded as a single-RU massive MIMO system.}
  \label{Fig:SysModel2}
\end{figure}

Consider a massive MIMO system as shown in Figs. {\ref{Fig:SysModel1}} and {\ref{Fig:SysModel2}}. One DU (cloud BS) controls three RUs and $K$ users. Each RU is connected with one another by optical fibers. Fig. {\ref{Fig:SysModel1}}(a) shows that the cloud BS provides a massive MIMO environment to cell-edge users under the assumption that RU has a relatively small number of antennas (more practical in the recent antenna configuration; we will use this system model in Section \ref{Section6}). Fig. {\ref{Fig:SysModel1}}(b) illustrates the equivalent model of Fig.~{\ref{Fig:SysModel1}}(a) considered as single-cell massive MIMO systems {\color{black}with cell edge users}. We assume that the cloud BS has $M$ antennas and each user equipment (UE) is equipped with one antenna. {\color{black}In this paper, we do not consider pilot contamination and assume perfect CSI at the RU.} We also assume that the channel is flat fading and the elements of a channel matrix are modeled as independent complex Gaussian random variables with zero mean and unit variance. The channel between the cloud BS (one DU and three RUs) and the $k$-th user is denoted by an $1 \times M$ row vector $\pmb{h}_{k}^T$ ($k=1, 2, \cdots, K$). A $K \times M$ channel matrix $\pmb{H}$ between the cloud BS and all UEs consists of channel vectors $\pmb{h}_{k}^T$. Let $\pmb{g}_k$ denote the column vector of transmit precoding and $s_k$ represent the transmit symbol for the $k$-th UE at downlink. Similarly, let $\pmb{w}_k$ denote the column vector of receive combining filter for the $k$-th UE at uplink. Also, let $n_k$ be the additive white Gaussian noise vector. Then, the received signal at the $k$-th UE is expressed by
\begin{align}
y_k=\underbrace{\sqrt{P_\text{t}}\pmb{h}_k^T\pmb{g}_ks_k}_{\text{desired signal}}+\underbrace{\sum_{\ell=1,\ell\neq k}^{K}\sqrt{P_\text{t}}\pmb{h}_k^T\pmb{g}_{\ell}s_\ell}_{\text{interference}}+n_k\label{eq1_1}
\end{align}
where, $P_\text{t}$ denotes the total network transmit power across three RUs. Also, the received signal for the $k$-th UE at the cloud BS is expressed by
\begin{align}
r_k=\underbrace{\sqrt{P_\text{u}}\pmb{w}_k^T\pmb{h}_kx_k}_{\text{desired signal}}+\underbrace{\sum_{\ell=1,\ell\neq k}^{K}\sqrt{P_\text{u}}\pmb{w}_k^T\pmb{h}_{\ell}x_\ell}_{\text{interference}}+\pmb{w}_k^Tn_k\label{eq1_2}
\end{align}
where, $P_\text{u}$ and $x_k$ denote the transmit power per each user and the transmit symbol of the $k$-th user at uplink, respectively.
\subsection{{\color{black}Downlink}}
Eq. (\ref{eq1_1}) contains the desired signal, interference, and noise terms. To eliminate the interference term, we use the following precoding:
\begin{align}
&\text{ZF} : \pmb{F}=\pmb{H}^*(\pmb{H}\pmb{H}^*)^{-1}=[\pmb{f}_1~ \pmb{f}_2~ \cdots~ \pmb{f}_k \cdots~ \pmb{f}_K],\nonumber\\
&\text{MRT} : \pmb{F}=\pmb{H}^*=[\pmb{f}_1~ \pmb{f}_2~ \cdots~ \pmb{f}_k \cdots~ \pmb{f}_K]\nonumber
\end{align}
where $\pmb{F}$ is a precoding matrix consisting of each column vector $\pmb{f}_k$.

{\color{black}To satisfy the power constraint, we consider two methods, i.e., vector/matrix normalizations.} The normalized transmit beamforming vectors (columns of a precoding matrix) with vector/matrix normalizations are given as $\pmb{g}_k=\pmb{f}_k/(\sqrt{K}||\pmb{f}_k||)$ and $\pmb{g}_k=\pmb{f}_k/||\pmb{F}||_F$, respectively. {\color{black}Note that vector normalization imposes equal power per downlink stream, while matrix normalization yields streams with different power. In this paper, to simplify, we do not consider a power optimization that could yield a complexity problem in very large array antenna systems.} 

\subsubsection{ZF/MRT with vector normalization}
The received signal at the $k$-th UE can be expressed as follows:
\begin{align}
y_k=\sqrt{P_\text{t}}\pmb{h}_k^T\frac{\pmb{f}_k}{\sqrt{K}||\pmb{f}_k||}s_k+\sum_{\ell=1,\ell\neq k}^{K}\sqrt{P_\text{t}}\pmb{h}_k^T\frac{\pmb{f}_{\ell}}{\sqrt{K}||\pmb{f}_{\ell}||}s_\ell+n_k.\label{ZFMRT vector norm. eq}
\end{align}

\subsubsection{ZF/MRT with matrix normalization}
Similarly, we can rewrite the received signal with matrix normalization as such:
\begin{align}
y_k=\sqrt{P_\text{t}}\pmb{h}_k^T\frac{\pmb{f}_k}{||\pmb{F}||_F}s_k+\sum_{\ell=1,\ell\neq k}^{K}\sqrt{P_\text{t}}\pmb{h}_k^T\frac{\pmb{f}_{\ell}}{||\pmb{F}||_F}s_\ell+n_k.\label{ZFMRT matrix norm. eq}
\end{align}

\subsection{{\color{black}Uplink}}
Similar to the downlink system, to eliminate the interference term, and to maximize the SNR in (\ref{eq1_2}), we use the following combining filter at the RUs:
\begin{align}
&\text{ZF} : \pmb{W}=(\pmb{H}^*\pmb{H})^{-1}\pmb{H}^*=[\pmb{w}_1~ \pmb{w}_2~ \cdots~ \pmb{w}_k \cdots~ \pmb{w}_K],\nonumber\\
&\text{MRC} : \pmb{W}=\pmb{H}^*=[\pmb{w}_1~ \pmb{w}_2~ \cdots~ \pmb{w}_k \cdots~ \pmb{w}_K]\nonumber
\end{align}
where $\pmb{W}$ is a combining filter matrix consisting of each column vector $\pmb{w}_k$. Here, we do not consider normalization since it does not change SNR values in the uplink scenario.

{\color{black}
\subsection{{\color{black}Summary of the Main Results}}
We first summarize the main results based on our analyses. These results are mathematically motivated from some random matrix theorems, {\color{black}shown} in Section \ref{SectionPre}.
\subsubsection{Asymptotic downlink sum rate}
In Section \ref{Section3}, we derive the simple and tight approximations of the ergodic achievable sum rate of ZF and MRT. We also find that ZF with vector normalization is better than ZF with matrix normalization while MRT with matrix normalization is better than MRT with vector normalization. The proposed ergodic achievable sum {\color{black}rates} of ZF with vector normalization at low SNR and MRT with matrix normalization at low SNR are given by
\begin{align}
\mathcal{R}_{\text{ZF}_{\text{vec}}, \text{ DL}_\text{L}}=K\log_2\left\{1+\frac{P_\text{t}(M-K+1)}{K}\right\},\nonumber\\
\mathcal{R}_{\text{MRT}_{\text{mat}}, \text{ DL}_{\text{L}}}\approx K\log_2\left\{1+\frac{P_\text{t}(M+1)}{P_\text{t}(K-1)+K}\right\}.\nonumber
\end{align}
\subsubsection{Asymptotic uplink sum rate}
We also evaluate the approximations of the ergodic achievable sum rate of ZF and MRC in the uplink case. The proposed ergodic achievable sum {\color{black}rates} of ZF at low SNR and MRC at low SNR are given by
\begin{align}
\mathcal{R}_{\text{MRC, UL}_{\text{L}}}\approx K\log_2\left\{1+\frac{P_\text{u}M}{P_\text{u}(K-1)+1}\right\},\nonumber\\
\mathcal{R}_{\text{ZF, UL}_{\text{L}}}\approx K\log_2\left\{1+P_\text{u}(M-K+1)\right\}.\nonumber
\end{align}
\subsubsection{Transceiver mode selection algorithm}
In Section \ref{Section5}, we propose two transceiver mode selection algorithms from i) {\color{black}a} transmit power and ii) {\color{black}the number of active users perspectives.} We explain how much transmit power and/or {\color{black}the} number of active users are needed for ZF to provide a better sum rate than MRT/MRC. The thresholds are given in \emph{Lemma}s~\ref{lemma1}-\ref{lemma5}.
}
{\color{black}\section{Mathematical Motivations and Preliminaries}\label{SectionPre}}
In this section, we introduce some mathematical {\color{black}motivations} and preliminaries to evaluate an asymptotic analysis for network massive MIMO, which will be used in Sections \ref{Section3} and \ref{Section4}.

\subsection{{\color{black}Achievable Rate Bound}}
In this paper, to maximize the achievable sum rate of downlink/uplink systems, we evaluate the closed forms of each system's performance. {\color{black}The achievable rates are bounded as follows:}
\begin{align}
\log_2\left\{1+\frac{1}{\mathbb{E}\left(\frac{I+N}{S}\right)}\right\}&\le\mathbb{E}\left\{\log_2\left(1+\frac{S}{I+N}\right)\right\}\nonumber\\
&\le \log_2\left\{1+\mathbb{E}\left(\frac{S}{I+N}\right)\right\}\nonumber
\end{align}
by using \emph{Jensen's Inequality} of convex and concave functions {\color{black}where $S$, $I$, and $N$ represent signal power, interference power, and noise power, respectively. Note that we use these bounds only for ZF cases and show that our results, based on the bounds, are more accurate than prior work \cite{Ngo_TCOM12}, \cite{Rusek_SPMag_12}, \cite{Jakob}.}

\subsection{Expectation and Variance of Random Vectors}
\begin{lemma}
Let $\pmb{h}_k$ and $\pmb{h}_\ell$ ($k\neq \ell$) be $M \times 1$ vectors whose elements are independent identically distributed (i.i.d.) complex Gaussian random variables with zero mean and unit variance.

{\color{black}
1) $\mathbb{E}[||\pmb{h}_k||^2]=M$,

~~~$\text{Var}[||\pmb{h}_k||^2]=M$.

2) $\mathbb{E}[\pmb{h}_k^*\pmb{h}_\ell]=0$,

~~~$\text{Var}[\pmb{h}_k^*\pmb{h}_\ell]=M$.

3) $\mathbb{E}[||\pmb{h}_k||^4]=M^2+M$,

~~~$\text{Var}[||\pmb{h}_k||^4]=4M^3+10M^2+6M$.

4) $\mathbb{E}[|\pmb{h}_k^*\pmb{h}_\ell|^2]=M$,

~~~$\text{Var}[|\pmb{h}_k^*\pmb{h}_\ell|^2]=M^2+2M$.
}
\label{pre1}
\begin{IEEEproof}
See Appendix \ref{Appendix Pre}.
\end{IEEEproof}
\end{lemma}

\subsection{Effective Channel}
\begin{lemma}
In massive MIMO systems, the transmit energy can be conserved by power-scaling law $1/M$ with perfect CSI.

{\color{black}
1) $\lim\limits_{M\to\infty}\mathbb{E}[\frac{1}{M}||\pmb{h}_k||^2]=1$,

~~~$\lim\limits_{M\to\infty}\text{Var}[\frac{1}{M}||\pmb{h}_k||^2]=0$.

2) $\lim\limits_{M\to\infty}\mathbb{E}[\frac{1}{M}\pmb{h}_k^*\pmb{h}_\ell]=0$,

~~~$\lim\limits_{M\to\infty}\text{Var}[\frac{1}{M}\pmb{h}_k^*\pmb{h}_\ell]=0$.
}
\label{pre2}
\end{lemma}

Generally, a transceiver uses MRT or MRC in massive MIMO, which means that the effective channel of the desired signal becomes one and the interference signal becomes zero as the number of antennas ($M$) goes to infinity, as illustrated:
{\color{black}
\begin{align}
\frac{1}{M}\pmb{H}\pmb{H}^*\xrightarrow{a.s.}\pmb{I}_K, ~\text{as}~ M\to\infty\nonumber.
\end{align}
}
\subsection{Signal and Interference Power}
\begin{lemma}
In a similar way, the expectation and the variance of the signal and the interference power are given by

{\color{black}
1) $\lim\limits_{M\to\infty}\mathbb{E}[\frac{1}{M^2}||\pmb{h}_k||^4]=1$,

~~~$\lim\limits_{M\to\infty}\text{Var}[\frac{1}{M^2}||\pmb{h}_k||^4]=0$.

2) $\lim\limits_{M\to\infty}\mathbb{E}[\frac{1}{M^2}|\pmb{h}_k^*\pmb{h}_\ell|^2]=0$,

~~~$\lim\limits_{M\to\infty}\text{Var}[\frac{1}{M^2}|\pmb{h}_k^*\pmb{h}_\ell|^2]=0$.
}
\label{pre3}
\end{lemma}
Note that {\color{black}the terms $\frac{1}{M}||\pmb{h}_k||^4$ and $\frac{1}{M}|\pmb{h}_k^*\pmb{h}_\ell|^2$ do not converge to $M$ and zero, respectively,} as $M$ goes to infinity since their variance does not go to zero, which means that they still have randomness. 
\subsection{Chebyshev's Inequality}\label{Cheby}
{\color{black}Let $X$ be a random variable with variance $\sigma_X^2$, $c$ and $\epsilon$ be scalars, and $Y=\frac{1}{c}X$ be a random variable with variance $\sigma_Y^2=\frac{1}{c^2}\sigma_X^2$, respectively. If $c$ is not very large in comparison with $\mathbb{E}\{X\}$, but $c^2>>\sigma_X^2$, then $Y=\frac{1}{c}X$ is {\emph{near deterministic}}, i.e., $P[|Y-\mathbb{E}\{Y\}|>\epsilon]\le\frac{\sigma_Y^2}{\epsilon^2}=\frac{\sigma_X^2}{\epsilon^2c^2}$. For fixed $\epsilon>0$, 
\begin{align}\label{pre4}
\text{if}~~\frac{\sigma_X^2}{c^2}\to0,~~\text{then}~~Y\approx\mathbb{E}\{Y\}
\end{align}
with high probability. We use (\ref{pre4}) to approximate a SINR term in ergodic sum rate expressions.

As an example, to analyze the rate in the high SNR regime, if $\text{Var}[\frac{1}{P_\text{t}}||\pmb{h}_k||^2]=\frac{1}{P_{\text{t}}^2}M$ converges to zero, then $\frac{1}{P_\text{t}}||\pmb{h}_k||^2$ converges to $\frac{1}{P_\text{t}}\mathbb{E}\{|\pmb{h}_k|^2\}=\frac{1}{P_\text{t}}M$. 

\subsection{Ergodic Achievable Sum Rate of Massive MIMO Systems {\color{black}at the low/high SNR regime}}

\begin{lemma}
Let $X_{\pmb{v}}$ and $v_i$ be a norm of a random vector $\pmb{v}$ ($M\times1$) and the $i$-th entry of $\pmb{v}$, respectively, i.e., $X_{\pmb{v}}=v_1^2+v_2^2+\dots+v_M^2$. {\color{black}Since $\mathbb{E}\left\{\frac{1}{X_{\pmb{v}}}\right\}=\mathbb{E}\left\{\frac{1}{v_1^2+v_2^2+\dots+v_M^2}\right\}=\mathbb{E}\left\{\frac{1}{M(v_1^2+v_2^2+\dots+v_M^2)/M}\right\}\approx\frac{1}{M\mathbb{E}\left\{v_i^2\right\}}$} and $\frac{1}{\mathbb{E}\left\{X_{\pmb{v}}\right\}}=\frac{1}{\mathbb{E}\left\{v_1^2+v_2^2+\dots+v_M^2\right\}}\approx\frac{1}{M\mathbb{E}\left\{v_i^2\right\}}$, $\mathbb{E}\left\{\frac{1}{X_{\pmb{v}}}\right\}$ converges to $\frac{1}{\mathbb{E}\left\{X_{\pmb{v}}\right\}}$ as {\color{black}$M$ goes to} infinity. 

In this {\emph{Lemma}}, {\color{black}we assume that the desired signal and the interference plus noise terms are norms of a random vector ($M\times1$), and some of those terms in the low/high SNR regime are \emph{near deterministic} as satisfying the condition of (\ref{pre4}); thus we assume at least one of the those terms has the same property of $X_{{\pmb{v}}}$.}
{\color{black}From the assumption,} if {\color{black}$S\approx\mathbb{E}\left\{S\right\}$} from (\ref{pre4}) with a low/high SNR assumption, then $\mathbb{E}\left\{\frac{S}{I+N}\right\}\approx\mathbb{E}\left\{S\right\}\mathbb{E}\left\{\frac{1}{I+N}\right\}$. Also, $\mathbb{E}\left\{S\right\}\mathbb{E}\left\{\frac{1}{I+N}\right\}\approx\frac{\mathbb{E}\left\{S\right\}}{\mathbb{E}\left\{I+N\right\}}$ as $M$ goes {\color{black}to} infinity using $\mathbb{E}\left\{\frac{1}{X_{\pmb{v}}}\right\}\approx\frac{1}{\mathbb{E}\left\{X_{\pmb{v}}\right\}}$. Similarly, for $I+N$ converges to $\mathbb{E}\left\{I+N\right\}$ cases, we also obtain $\mathbb{E}\left\{\frac{S}{I+N}\right\}\approx\frac{\mathbb{E}\left\{S\right\}}{\mathbb{E}\left\{I+N\right\}}$. So we could obtain the following approximation of SINR when {\color{black}\emph{$M$ goes to infinity {\color{black}in} the low or high SNR regime}}:
\begin{align}
\mathbb{E}\left\{\frac{S}{I+N}\right\}\approx\frac{\mathbb{E}(S)}{\mathbb{E}(I+N)}.
\label{approx SNR}
\end{align}

From (\ref{approx SNR}), the lower bound of the ergodic sum rate is the same as the upper bound of the ergodic sum rate. Thus we could also get the following approximation of the ergodic sum rate when {\color{black}\emph{$M$ goes to infinity {\color{black}in} the low or high SNR regime}}: 
\begin{align}
\mathbb{E}\left(\log_2\left(1+\frac{S}{I+N}\right)\right)\approx\log_2\left(1+\frac{\mathbb{E}(S)}{\mathbb{E}(I+N)}\right).\nonumber
\end{align}
\label{pre5}
\end{lemma}

\subsection{Arithmetic-Geometric Inequality}

%
\begin{lemma} 
Let $b_1, b_2, ..., b_K$ be random variables. We can obtain the following inequality through \emph{Arithmetic-geometric Inequality} defined in \cite{Gradshteyn_Table}:
$$
\sum_{k=1}^K\log_2\left(1+\frac{1}{Kb_k}\right)\ge K\log_2\left(1+\frac{1}{\sum_{k=1}^Kb_k}\right).
$$
\begin{IEEEproof}
\begin{align}
&\frac{1}{K}\sum_{k=1}^K\log_2\left(1+\frac{1}{Kb_k}\right)\ge \log_2\left(1+\frac{1}{\frac{1}{K}\sum_{k=1}^KKb_k}\right)\nonumber\\
&\Leftrightarrow\sum_{k=1}^K\log_2\left(1+\frac{1}{Kb_k}\right)\ge K\log_2\left(1+\frac{1}{\sum_{k=1}^Kb_k}\right).\nonumber
\end{align}
\end{IEEEproof}
\label{pre6}
\end{lemma}
This is a simple application of \emph{Jensen's Inequality}.
\section{{\color{black}Asymptotic Downlink Sum Rate for Cell-boundary Users}}\label{Section3}
In this section, we derive the achievable rate bounds, and show which normalization method is suitable for ZF- and MRT-type precoding at the downlink. 
{\color{black}We assume that the cell-boundary users are in the low SNR regime ($P_\text{t}\text{Var}\left\{X_v\right\}\to0$) and that the cell center users are in the high SNR regime ($\frac{1}{P_\text{t}}\text{Var}\left\{X_v\right\}\to0$); we also assume that the desired power term or the interference power term becomes \emph{near deterministic} in the low/high SNR regime.} Based on our analytical results, we will also show which precoding technique is desired for cell-boundary users. 

\subsection{Ergodic Performance}
\subsubsection{Achievable rate bounds for ZF precoding}\

~i) Lower bounds:
The lower bound of the ergodic sum rate for the ZF precoding is well known, as follows~\cite{Rusek_SPMag_12}:
\begin{align}
\begin{split}
&\mathcal{R}_{\text{ZF}_{\text{vec}}, \text{ DL}}^L=\mathcal{R}_{\text{ZF}_{\text{mat}}, \text{ DL}}^L=K\log_2\left\{1+\frac{P_\text{t}(M-K)}{K}\right\}\nonumber
\end{split}
\end{align}
using the property {\color{black}of Wishart matrices}~\cite{RandomMatrixBook}. 

~ii) {Vector normalization-upper bound}:
From (\ref{ZFMRT vector norm. eq}), we can derive the SINR of the upper bound of vector normalization in the ZF case, {\color{black}as given by}
\begin{align}
\mathbb{E}\left\{\frac{S}{I+N}\right\}&=\mathbb{E}\left\{\frac
{P_\text{t}\left|\pmb{h}_k^T\frac{\pmb{f}_k}{\sqrt{K}||\pmb{f}_k||}\right|^2}{P_\text{t}\sum_{\ell=1,\ell\neq k}^{K}\left|\pmb{h}_k^T\frac{\pmb{f}_{\ell}}{\sqrt{K}||\pmb{f}_{\ell}||}\right|^2+1}\right\}\nonumber\\
&=\mathbb{E}\left\{\frac{P_\text{t}}{K||\pmb{f}_k||^2}\right\}\nonumber\\
&\overset{(a)}{=}\frac{P_\text{t}(M-K+1)}{K}\label{ZF VectorNorm Up eq}
\end{align}
where {\color{black}$(a)$} results from the diversity order of ZF {\color{black}$\left(\mathbb{E}\left\{\frac{1}{||\pmb{f}_k||^2}\right\}=M-K+1\right)$}, as shown in \cite{Wong_08}. From (\ref{ZF VectorNorm Up eq}), the upper bound of vector normalization in the ZF case can be represented as
\begin{align}
\begin{split}
&\mathcal{R}_{\text{ZF}_{\text{vec}}, \text{ DL}}^U=K\log_2\left\{1+\frac{P_\text{t}(M-K+1)}{K}\right\}.\nonumber
\end{split}
\end{align}

~iii) {Matrix normalization-upper bound}:
From (\ref{ZFMRT matrix norm. eq}), the SINR of the upper bound of matrix normalization in the ZF case can be expressed as
\begin{align}
\mathbb{E}\left\{\frac{S}{I+N}\right\}&=\mathbb{E}\left\{\frac{P_\text{t}\left|\pmb{h}_k^T\frac{\pmb{f}_k}{||\pmb{F}||_F}\right|^2}
{P_\text{t}\sum_{\ell=1,\ell\neq k}^{K}\left|\pmb{h}_k^T\frac{\pmb{f}_{\ell}}{||\pmb{F}||_F}\right|^2+1}\right\}\nonumber\\
&=\mathbb{E}\left\{P_\text{t}\left|\frac{1}{||\pmb{F}||_F}\right|^2\right\}.
\label{ZF MatrixNorm Upp eq1}
\end{align}
From (\ref{ZF MatrixNorm Upp eq1}), the first upper bound (with an expectation form) of matrix normalization in the ZF case can be represented as
\begin{align}
&\mathcal{R}_{\text{ZF}_{\text{mat}}, \text{ DL}}^{U1}=K\log_2\left\{1+\mathbb{E}\left(P_\text{t}\left|\frac{1}{||\pmb{F}||_F}\right|^2\right)\right\}.
\label{RZF mat up1}
\end{align}
By using \emph{Lemma} \ref{pre6}, (\ref{RZF mat up1}) can further be expressed as
\begin{align}
\begin{split}
&K\log_2\left\{1+\mathbb{E}\left(P_\text{t}\left|\frac{1}{||\pmb{F}||_F}\right|^2\right)\right\}\nonumber\\
&=K\log_2\left\{1+\mathbb{E}\left(P_\text{t}\frac{1}{\text{tr}((\pmb{H}\pmb{H}^*)^{-1})}\right)\right\}\nonumber\\
&=K\log_2\left\{1+\mathbb{E}\left(P_\text{t}\frac{1}{\sum_{k=1}^K||\pmb{f}_k||^2}\right)\right\}\nonumber\\
&\le K\log_2\left\{1+\mathbb{E}\left(P_\text{t}\frac{1}{K||\pmb{f}_k||^2}\right)\right\}.
\end{split}
\end{align}
So the second upper bound (without an expectation form) of matrix normalization in the ZF case can be given by
\begin{align}
&\mathcal{R}_{\text{ZF}_{\text{mat}}, \text{ DL}}^{U2}=K\log_2\left\{1+\frac{P_\text{t}(M-K+1)}{K}\right\}.
\label{RZF mat up2}
\end{align}

\subsubsection{Achievable rate for MRT precoding}\

~i) {Vector normalization-low SNR regime}:
From (\ref{ZFMRT vector norm. eq}), we can derive the ergodic achievable sum rate of vector normalization in low SNR as follows:
\begin{align}
\mathcal{R}_{\text{MRT}_{\text{vec}}, \text{ DL}_{\text{L}}}
&\approx K\log_2\left\{1+\frac{P_\text{t}M}{P_\text{t}(K-1)+K}\right\}.
\label{MRT vector low SNR}
\end{align}
\begin{IEEEproof}
See Appendix \ref{Appendix B1}.
\end{IEEEproof}
~ii) {Vector normalization-high SNR regime}:
Similarly, we can get the ergodic achievable sum rate of vector normalization in high SNR as follows:
\begin{align}
\mathcal{R}_{\text{MRT}_{\text{vec}}, \text{ DL}_{\text{H}}}
&\approx K\log_2\left\{1+\frac{P_\text{t}(M+1)}{P_\text{t}(K-1)+K}\right\}.
\label{MRT vector high SNR}
\end{align}
\begin{IEEEproof}
See Appendix \ref{Appendix B2}.
\end{IEEEproof}
~iii) {Matrix normalization-low/high SNR regime}:
From (\ref{ZFMRT matrix norm. eq}), we can evaluate the ergodic achievable sum rate of matrix normalization in low/high SNR by using the following formation:
\begin{align}
&\mathcal{R}_{\text{MRT}_{\text{mat}}, \text{ DL}_{\text{L/H}}}\approx K\log_2\left\{1+\frac{P_\text{t}(M+1)}{P_\text{t}(K-1)+K}\right\}.
\label{MRT matrix low high SNR}
\end{align}
\begin{IEEEproof}
See Appendix \ref{Appendix B3} and \ref{Appendix B4}.
\end{IEEEproof}

\subsection{Comparison between Vector and Matrix Normalizations}\label{BestNormSection}
\subsubsection{Performance comparison of ZF}
To find which normalization technique is better in ZF, we let $b_k=\frac{[(\pmb{H}\pmb{H}^*)^{-1}]_{kk}}{P_\text{t}}$ in \emph{Lemma} \ref{pre6} directly.
\begin{align}
\sum_{k=1}^K\log_2\left(1+\frac{P_\text{t}}{K[(\pmb{H}\pmb{H}^*)^{-1}]_{kk}}\right) ~~~~~~~~~~~~~~\nonumber\\
~~~~~~~~~~~~\ge K\log_2\left(1+\frac{P_\text{t}}{\sum_{k=1}^K[(\pmb{H}\pmb{H}^*)^{-1}]_{kk}}\right)\nonumber\\
\Leftrightarrow\mathcal{R}_{\text{ZF}_{\text{vec}}, \text{ DL}}\ge\mathcal{R}_{\text{ZF}_{\text{mat}}, \text{ DL}}~~~~~~~~~~~~~~~
\label{comparisonZF}
\end{align}
{\color{black}where $[(\pmb{H}\pmb{H}^*)^{-1}]_{kk}=||\pmb{f}_k||^2$ and $\sum_{k=1}^K[(\pmb{H}\pmb{H}^*)^{-1}]_{kk}=\sum_{k=1}^K\||\pmb{f}_k||^2=||\pmb{F}||^2$. From (\ref{comparisonZF}), since the desired power and the interference power are one and zero respectively, the power normalization per user only affects the performance of ZF. From the perspective of \emph{Jesen's Inequality}, the sum rate with the different power allocation per user is the upper bound of the sum rate with the same power allocation per user at the instant channel and arbitrary SNR. We can conclude that, in the ZF case, vector normalization is always better than matrix normalization.}

\subsubsection{Performance comparison of MRT}
From (\ref{MRT vector low SNR})-(\ref{MRT matrix low high SNR}), a comparison of the ergodic achievable sum rate is given by
\begin{align}
\mathcal{R}_{\text{MRT}_{\text{mat}}, \text{ DL}_{\text{L}}}\gtrapprox\mathcal{R}_{\text{MRT}_{\text{vec}}, \text{ DL}_{\text{L}}}\label{comparisonMRT low}\\
\mathcal{R}_{\text{MRT}_{\text{mat}}, \text{ DL}_{\text{H}}}\approx\mathcal{R}_{\text{MRT}_{\text{vec}}, \text{ DL}_{\text{H}}}\label{comparisonMRT high}
\end{align}
at low and high SNR, respectively. 
{\color{black}It is well known that approach methods that maximize the desired signal are better than those that mitigate the interference power at low SNR. The effective desired channel gain can be maximized with MRT. The desired power per user scales down in proportion to its power in the vector normalization while the desired power per user scales down in the same proportion over all users. This means that the gains of better channels with vector normalization tend to scale down more than the gains of better channels with matrix normalization. This yields the difference of the desired power term of (\ref{comparisonMRT low}) and (\ref{comparisonMRT high}) at low SNR.} 
Therefore, we confirm that, for MRT precoding, matrix normalization is always better than vector normalization at low SNR. We conclude, however, that there is marginal performance gap between vector normalization and matrix normalization at high SNR.

\subsection{Ergodic Achievable Sum Rate for Cell-boundary Users with the Best Normalization Method}
As explained {\color{black}in Section~\ref{BestNormSection}}, we conclude that the suitable normalization methods are vector normalization for ZF and matrix normalization for MRT. We assume that the transmit power ($P_\text{t}$) is small for cell-boundary users (low SNR regime). Using the property of ZF precoding, the ergodic achievable sum rate of ZF is represented as{\color{black}
\begin{align}
&\mathbb{E}\left\{\log_2\left(1+\frac{S}{I+N}\right)\right\}\nonumber\\
&=\mathbb{E}\left\{\log_2\left(1+\frac{P_\text{t}|\pmb{h}_k^T\pmb{g}_k|^2}{P_\text{t}\sum_{\ell=1,\ell\neq k}^{K}|\pmb{h}_k^T\pmb{g}_\ell|^2+1}\right)\right\}\nonumber\\
&=\mathbb{E}\left\{\log_2\left(1+{P_\text{t}|\pmb{h}_k^T\pmb{g}_k|^2}\right)\right\}\nonumber\\
&\overset{(b)}{\approx}\log_2\left\{1+P_\text{t}\mathbb{E}(|\pmb{h}_k^T\pmb{g}_k|^2)\right\}
\label{ZF_low}
\end{align}
where ($b$)} results from (\ref{pre4}) {\color{black}with $P_\text{t}\text{Var}\left\{|\pmb{h}_k^T\pmb{g}_k|^2\right\}\to0$}. Eq.~(\ref{ZF_low}) indicates that the achievable sum rate of ZF precoding can approach its upper bounds at low SNR by (\ref{pre4}). Thus, the ergodic achievable sum rate of ZF with vector normalization at low SNR is given by{\color{black}
\begin{align}
&\mathcal{R}_{\text{ZF}_{\text{vec}}, \text{ DL}_\text{L}}\approx K\log_2\left\{1+\frac{P_\text{t}(M-K+1)}{K}\right\}.\nonumber
\end{align}}
From (\ref{MRT matrix low high SNR}), we find the ergodic achievable sum rate of matrix normalization in low SNR:
\begin{align}
&\mathcal{R}_{\text{MRT}_{\text{mat}}, \text{ DL}_{\text{L}}}\approx K\log_2\left\{1+\frac{P_\text{t}(M+1)}{P_\text{t}(K-1)+K}\right\}\nonumber.
\end{align}




\section{Asymptotic Uplink Sum Rate for Cell-Boundary Users}\label{Section4}
We have focused on a downlink scenario with a sum power constraint. In this section, we investigate an uplink case, where each user has its own power constraint. From (\ref{eq1_2}), the ergodic achievable sum rate for the uplink, $\mathcal{R}_\text{UL}$, is 
\begin{align}
\mathcal{R}_\text{UL}=\mathbb{E}\left[\sum_{k=1}^{K}\log_2\left\{1+\frac{P_\text{u}|\pmb{w}_k^T\pmb{h}_k|^2}{P_\text{u}\sum_{\ell=1,\ell\neq k}^{K}|\pmb{w}_k^T\pmb{h}_\ell|^2+||\pmb{w}_k||^2}\right\}\right].\label{eq4_1}
\end{align}
From (\ref{eq4_1}), we can derive the ergodic achievable uplink sum rate of MRC, $\mathcal{R}_\text{MRC, UL}$, as follows: 
\begin{align}
&\mathcal{R}_\text{MRC, UL}\nonumber\\
&=\mathbb{E}\left[\sum_{k=1}^{K}\log_2\left\{1+\frac{P_\text{u}||\pmb{h}_k||^4}{P_\text{u}\sum_{\ell=1,\ell\neq k}^{K}|\pmb{h}_k^{*}\pmb{h}_\ell|^2+||\pmb{h}_k||^2}\right\}\right].\label{eq4_2}
\end{align}
We approximate the ergodic achievable sum rate of MRC as follows:

i) High SNR regime:
\begin{align}
&\mathcal{R}_{\text{MRC, UL}_{\text{H}}}\approx K\log_2\left\{1+\frac{P_\text{u}(M+1)}{P_\text{u}(K-1)+1}\right\}
\label{eq4_3}.
\end{align}

ii) Low SNR regime:
\begin{align}
&\mathcal{R}_{\text{MRC, UL}_{\text{L}}}\approx K\log_2\left\{1+\frac{P_\text{u}M}{P_\text{u}(K-1)+1}\right\}
\label{eq4_4}.
\end{align}
\begin{IEEEproof}
See Appendix \ref{Appendix C}
\end{IEEEproof}
{\color{black}Similar to ZF precoding, the ergodic sum rate for ZF for uplink at low SNR, $\mathcal{R}_{\text{ZF, UL}_{\text{L}}}$, is 
\begin{align}
\begin{split}
\mathcal{R}_{\text{ZF, UL}_{\text{L}}}\approx \mathcal{R}_{\text{ZF, UL}}^{U}&=K\log_2\left\{1+\mathbb{E}\left\{\frac{P_\text{u}}{[(\pmb{H}^{*}\pmb{H})^{-1}]_{k,k}}\right\}\right\}\nonumber\\
&=K\log_2\left\{1+P_\text{u}(M-K+1)\right\}.\nonumber
\end{split}
\end{align} }


\section{{\color{black}Transceiver Mode Selection}}\label{Section5}

\subsection{{\color{black}Algorithm}}

In this section, we propose two transceiver mode selection algorithms from {\color{black}i) transmit power and ii) the number of active users perspectives}. To provide a mathematically simple solution, we first {\color{black}introduce} \emph{Lemma}~\ref{lemma1} and \emph{Lemma}~\ref{lemma2} {\color{black}that} use a power threshold as follows:
 
\begin{lemma}
The power threshold to select a better precoder for downlink is given by
\begin{align}
P_\text{th, DL}=\frac{K^2}{(K-1)(M-K+1)}.\label{eq5_1}
\end{align} 

If the RUs have more transmit power than the power threshold $P_\text{th, DL}$, the ZF precoder provides a better sum rate performance. 
\begin{IEEEproof}
To derive (\ref{eq5_1}) for cell-boundary users, we use the low SNR approximation for ZF and MRT. By letting $\mathcal{R}_{\text{ZF}_{\text{vec}},\text{ DL}_\text{L}}\ge\mathcal{R}_{\text{MRT}_{\text{mat}},\text{ DL}_\text{L}}$, we can get (\ref{eq5_1}) as follows:
\begin{align}
\begin{split}
&~~~~\mathcal{R}_{\text{ZF}_\text{vec}\text{, DL}_\text{L}}-\mathcal{R}_{\text{MRT}_\text{mat}\text{, DL}_\text{L}}\ge0\nonumber\\
&\Leftrightarrow \frac{P_{\text{t}}(M-K+1)}{K}-\frac{P_{\text{t}}(M+1)}{P_{\text{t}}(K-1)+K}\ge0\nonumber\\
&\Leftrightarrow P_{\text{t}}\ge P_\text{th, DL}=\frac{K^2}{(K-1)(M-K+1)}.\nonumber
\end{split}
\end{align}\end{IEEEproof}
\label{lemma1}
\end{lemma} 
 
 \begin{lemma}
 {\color{black}The power threshold to select a better receive combining filter at uplink is given~by
\begin{align}
P_\text{th, UL}=\frac{1}{M-K+1}\label{eq5_3}.
\end{align}}
If each UE has larger transmit power per user than $P_\text{th, UL}$, the solution employing ZF at the RUs provides a better sum rate performance. 
\begin{IEEEproof}
To evaluate (\ref{eq5_3}), we use the low SNR approximation of MRC, i.e., (\ref{eq4_4}). From $\mathcal{R}_\text{ZF, UL}^L\ge\mathcal{R}_{\text{MRC, UL}_{\text{L}}}$, we can obtain (\ref{eq5_3}) for uplink as follows:
\begin{align}
\begin{split}
&~~~~\mathcal{R}_{\text{ZF, UL}_{\text{L}}}-\mathcal{R}_{\text{MRC, UL}_{\text{L}}}\ge0\nonumber\\
&\Leftrightarrow P_\text{u}(M-K+1)-\frac{P_{\text{u}}M}{P_{\text{u}}(K-1)+1}\ge0\nonumber\\
&\Leftrightarrow P_{\text{u}}\ge P_{\text{th ,UL}}=\frac{1}{M-K+1}.\nonumber
\end{split}
\end{align}
\end{IEEEproof}
\label{lemma2}
\end{lemma} 
 
\emph{Lemma} \ref{lemma1} helps the RUs select one of the precoders, i.e., ZF or MRT, with respect to the transmit power of the cloud BS. Also, the power policy of the cloud BS could be adjusted by the power threshold that is a function of $M$ and $K$. Therefore, the RUs could find a suitable precoding mode according to the user's {\color{black}location}. Similarly, \emph{Lemma} \ref{lemma2} could be applied to the uplink case. 

The proposed power threshold would be affected by a specific number of users, so a power cross point, that refers to $P_\text{cross}$, exists where MRT or MRC is always better for any number of active users. 
{\color{black}Since \emph{Lemma}s~\ref{lemma1} and \ref{lemma2} are monotonic increasing functions of $K$ ($P_\text{th, DL/UL}(K=k+1)>P_\text{th, DL/UL}(K=k)$), $P_\text{th, DL}$ and $P_\text{th, UL}$ have minimum values at $K=2$.} These points become $P_\text{cross, DL}$ and $P_\text{cross, UL}$.
\begin{lemma}
If the transmit power of the RUs/UEs is lower than $P_\text{cross}$, MRT or MRC is always better than ZF in terms of sum rate. The power cross point, $P_\text{cross}$, at downlink and uplink are given by
\begin{align}
P_\text{cross, DL}=\frac{4}{M-1}, ~~P_\text{cross, UL}=\frac{1}{M-1}\nonumber.
\end{align}
\label{lemma3}
\end{lemma}

Now we investigate $K_\text{cross}$, which is a transceiver mode selection threshold when the transmit power at the transceiver is larger than $P_\text{cross}$ and when the number of active users varies.

\begin{lemma} 
If the RUs have more transmit power than $P_\text{cross}$, the user cross point at downlink, $K_\text{cross, DL}$, for selecting a better precoder is given by
\begin{align}
K_\text{cross, DL}=\frac{P_\text{t}(M+1)}{1+P_\text{t}}.\label{eq5_5}
\end{align}
If the number of users $K$ is larger than $K_\text{cross, DL}$, MRT precoder provides a better sum rate performance.
\label{lemma4}
\end{lemma}

\begin{figure}[t]
 \centerline{\resizebox{1\columnwidth}{!}{\includegraphics{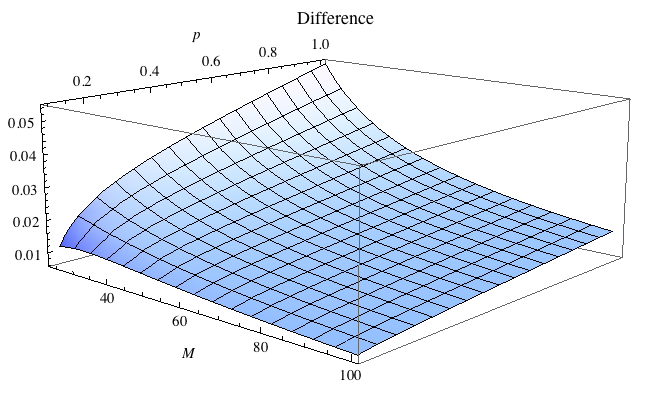}}}
  \caption{The difference of the gradient between ZF and MRT at $K_{\text{cross}}$ when $P$ is very small (almost zero) and $M$ is much larger than $P$. The difference is always positive ($>0$).}
  \label{Difference}
\end{figure}

\begin{IEEEproof}
Both $\mathcal{R}_{\text{ZF}_{\text{vec}},\text{ DL}_\text{L}}$ and $\mathcal{R}_{\text{MRT}_{\text{mat}},\text{ DL}_\text{L}}$ are concave functions. Also, unlike $\mathcal{R}_{\text{ZF}_{\text{vec}},\text{ DL}_\text{L}}$, $\mathcal{R}_{\text{MRT}_{\text{mat}},\text{ DL}_\text{L}}$ is a monotonic increasing function; thus, two cross points exist: one is when the number of users $K$ is one; the other is when the number of users $K$ has (\ref{eq5_5}) with a large $M$ approximation (this approximation results from satisfying $K_\text{cross, DL}=1$ condition) as follows:
\begin{align}
&~~~~\mathcal{R}_{\text{MRT}_\text{mat}\text{, DL}_\text{L}}-\mathcal{R}_{\text{ZF}_\text{vec}\text{, DL}_\text{L}}\ge0\nonumber\\
&\Leftrightarrow \frac{P_{\text{t}}(M+1)}{P_{\text{t}}(K-1)+K}-\frac{P_{\text{t}}(M-K+1)}{K}\nonumber\\
&~~~~\approx\frac{P_{\text{t}}M}{P_{\text{t}}(K-1)+K}-\frac{P_{\text{t}}(M-K+1)}{K}\ge0\nonumber\\
&\Leftrightarrow K\ge K_\text{cross, DL}=\frac{P_\text{t}(M+1)}{1+P_\text{t}}.\nonumber
\end{align}
\end{IEEEproof}
\begin{lemma}
{\color{black}The user cross point, $K_\text{cross, UL}$, to select a better receive combining filter at uplink when the RUs have larger transmit power than $P_\text{cross, UL}$, is given by
\begin{align}
K_\text{cross, UL}=M+1-\frac{1}{P_\text{u}}. \label{eq5_7}
\end{align}}
If the number of users, $K$, is larger than $K_{\text{cross, UL}}$, MRC provides a better sum rate performance.
\begin{IEEEproof}
Similar to \emph{Lemma} \ref{lemma4}, we can obtain (\ref{eq5_7}) as follows:
\begin{align}
\begin{split}
&~~~~\mathcal{R}_{\text{MRC, UL}_{\text{L}}}-\mathcal{R}_{\text{ZF, UL}_{\text{L}}}\ge0\nonumber\\
&\Leftrightarrow \frac{P_{\text{u}}M}{P_\text{u}(K-1)+1}-P_\text{u}(M-K+1)\nonumber\\
&\Leftrightarrow K\ge K_\text{cross, UL}=M+1-\frac{1}{P_\text{u}}.\nonumber
\end{split}
\end{align}
\end{IEEEproof}
\label{lemma5}
\end{lemma}

\emph{Lemma}s \ref{lemma3}-\ref{lemma5} provide a proper solution for the low SNR regime like a cell-boundary. For example, if the users have very low SNR, which means $P_\text{t}$ or $P_\text{u}$ is always lower than $P_\text{cross, DL}$ or $P_\text{cross, UL}$, the cloud BS should use MRT or MRC to increase a sum rate. Also, the cloud BS should use MRT or MRC for users having transmit power larger than $P_{\text{cross}}$ (especially in the low SNR regime) when the number of active users is larger than $K_\text{cross}$.
\begin{figure}[!t]
 \centering
    \vertfig[Achievable rate of ZF at low SNR.]{\label{ZF}\includegraphics[width=0.5\textwidth]{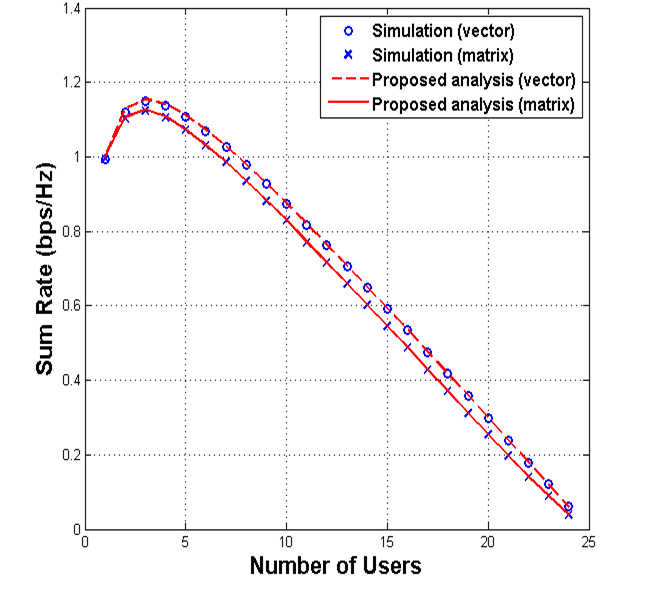}}
    \vertfig[Achievable rate of MRT at low SNR.]{\label{MRT}\includegraphics[width=0.5\textwidth]{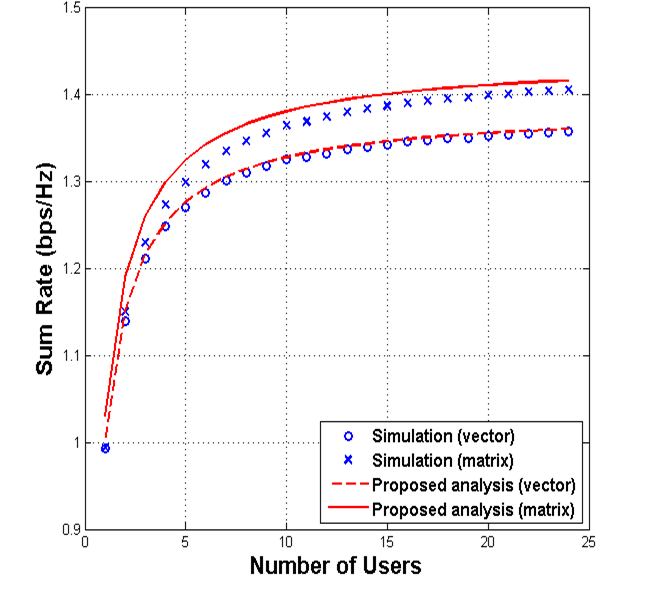}}
	\caption{Achievable rate vs. the number of cell-boundary users, where $M$ = 24, $K$ = [1, 24], and total SNR = $-13.8$~dB.}
\label{ZFMRT}
\end{figure}
\begin{figure}[!t]
 \centering
    \vertfig[Achievable rate of MRC at high SNR.]{\label{f3}\includegraphics[width=0.5\textwidth]{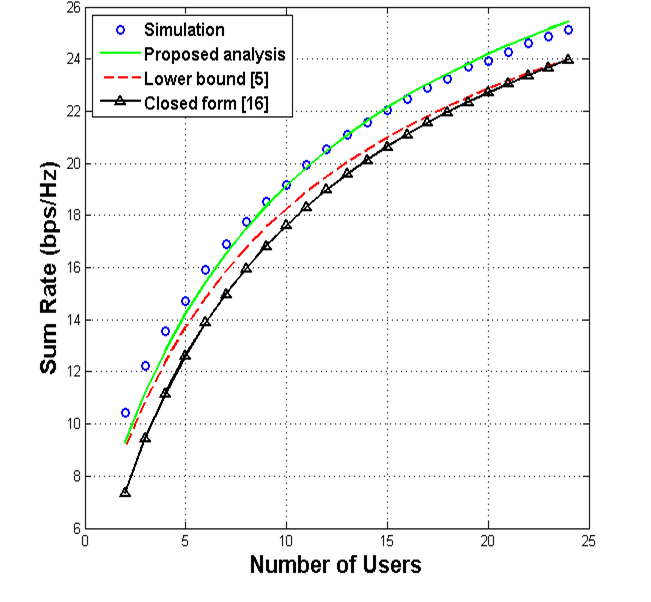}}
    \vertfig[Achievable rate of MRC at low SNR.]{\label{f4}\includegraphics[width=0.5\textwidth]{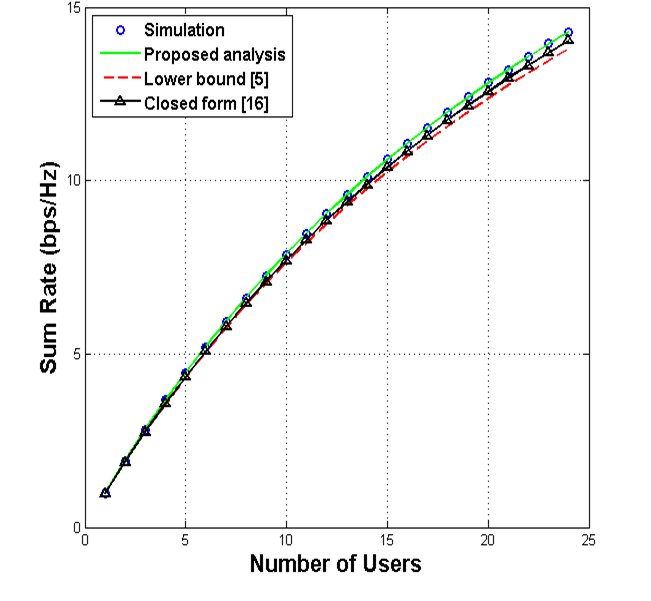}}
    \caption{Achievable rate vs. the number of cell-boundary users, where $M$ = 24, $K$ = [1, 24], and total SNR = (a) $13.8$~dB, (b) $-13.8$~dB.}
\label{f3f4}
\end{figure}

\subsection{{\color{black}Performance Comparison {\color{black}for} a Large Number of Users {\color{black}Case}}}
We derive the ergodic sum rate of ZF with vector normalization at low SNR (or upper bound of ZF with vector normalization) when $M=K$
\begin{align}
\lim\limits_{M\to\infty}\mathcal{R}_{\text{ZF}_{\text{vec}}, \text{ DL}_\text{L}}(M=K)
&=\lim\limits_{K\to\infty}K\log_2\left\{1+\frac{P_\text{t}}{K}\right\} \nonumber\\
&=\lim\limits_{K\to\infty}P_\text{t}\log_2e\ln\left\{1+\frac{P_\text{t}}{K}\right\}^{K/P_\text{t}}\nonumber\\
&=P_\text{t}\log_2e. \label{MK}
\end{align}
This is equal to the result in \cite{PerturPart1}. From this result, we are able to gather insights into user scheduling. 
{\color{black}We also derive the ergodic sum rate of MRT with any normalization and the ergodic sum rate of MRC when $M=K$
\begin{align}
\lim\limits_{M\to\infty}\mathcal{R}_\text{MRT, DL}(M=K)&=M\log_2\left\{1+\frac{P_\text{t}}{P_\text{t}+1}\right\}\nonumber\\
\lim\limits_{M\to\infty}\mathcal{R}_\text{MRC, UL}(M=K)&=M\log_2(1+1)=M\label{MF_largeK}.
\end{align}
In the special case of MRC, if the transmit power per user is scaling down with $M$ from the sum power constraint $P_\text{u, sum}$, i.e., $P_\text{u}=P_\text{u, sum}/M$, then
\begin{align}
\lim\limits_{M\to\infty}\mathcal{R}_\text{MRC, UL}(M=K)&=M\log_2\left\{1+\frac{P_\text{u, sum}}{P_\text{u, sum}+1}\right\}\label{MF_largeK_sc}.
\end{align}
From (\ref{MF_largeK}) and (\ref{MF_largeK_sc}), we conclude that the performance of MRT and MRC is bounded by $M$ when $M=K$, even SNR goes to infinity. This differs from the result in \cite{Ngo_TCOM12} ($\mathcal{R}_\text{MRC, UL}=M\log_2(1+P_\text{u, sum})$) because the authors in \cite{Ngo_TCOM12} did not consider the case of a large number of users.}

Next, at $K_{\text{cross, DL}}$ in downlink, we check the difference of the gradient between the rates of ZF and MRT. If the gradient of the rate of ZF is larger than that of MRT, the rate of ZF with vector normalization is larger than that of MRT when $K < K_{\text{cross, DL}}$. In the other case, the rate of MRT with matrix normalization is larger than that of ZF when $K \ge K_{\text{cross, DL}}$. The difference of the gradient between the rates of ZF and MRT is expressed as
\begin{align}
&~~~\mathcal{G}_{\text{MRT}_{\text{mat}},\text{ DL}_\text{L}}-\mathcal{G}_{\text{ZF}_{\text{vec}},\text{ DL}_\text{L}}\nonumber \\
 &=\frac{(P_\text{t}+1)^2}{(M+1)P\ln{4}}-\frac{(M+1)(P_\text{t}+1)}{MP\ln{2^{(2M+1)}}}\label{Diff_gradient}
\end{align}
where $\mathcal{G}_{\text{MRT}_{\text{mat}},\text{ DL}_\text{L}}$ denotes the gradient of the $\mathcal{R}_{\text{MRT}_{\text{mat}},\text{ DL}_\text{L}}$ curve at $K_{\text{cross, DL}}$. Similarly, $\mathcal{G}_{\text{ZF}_{\text{vec}},\text{ DL}_\text{L}}$ is the gradient of the $\mathcal{R}_{\text{ZF}_{\text{vec}},\text{ DL}}^L$ curve at $K_{\text{cross, DL}}$. In general, cell-boundary users have relatively low SNR and, as we assumed, the cloud BS has large-scale antennas, meaning $M$ is much larger than $P_\text{t}$. Therefore, if $K_{\text{cross, DL}}$ exists, (\ref{Diff_gradient}) is always positive. We also confirm this through numerical comparisons as shown in~Fig.~\ref{Difference}. From this observation, we realize that MRT precoding is suitable for cell-boundary users if the number of active users is larger than $K_{\text{cross, DL}}$.

\section{Numerical results}\label{Section6}

\begin{figure}[!t]
 \centering
    \vertfig[Achievable rate at downlink.]{\label{a}\includegraphics[width=0.5\textwidth]{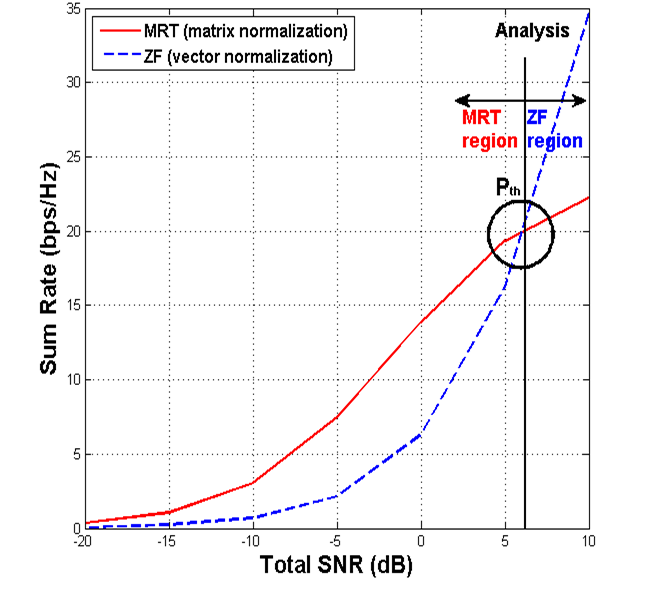}}
    \vertfig[Achievable rate at uplink.]{\label{b}\includegraphics[width=0.5\textwidth]{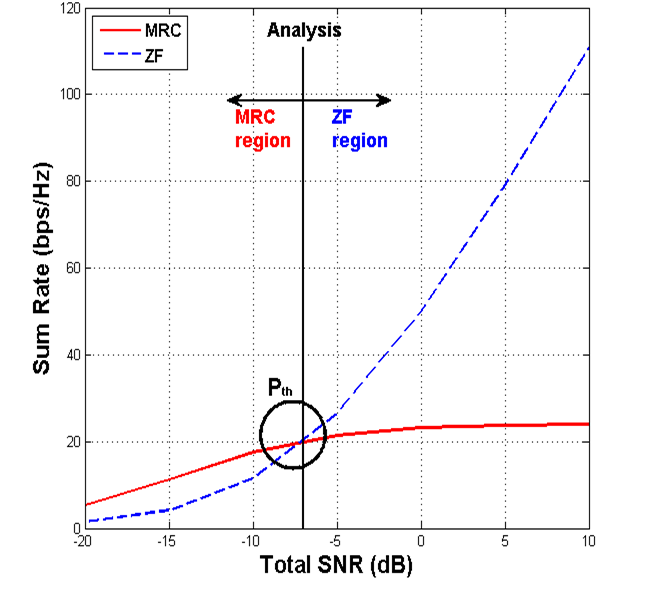}}
    \caption{Achievable rate vs total SNR, where $M$ = 24, $K$ = 20, and $P_\text{th}$= (a) $6.2$~dB, (b) $-7$~dB.}
\label{ab}
\end{figure}
\begin{figure}[!t]
 \centering
    \vertfig[Achievable rate at downlink.]{\label{c}\includegraphics[width=0.5\textwidth]{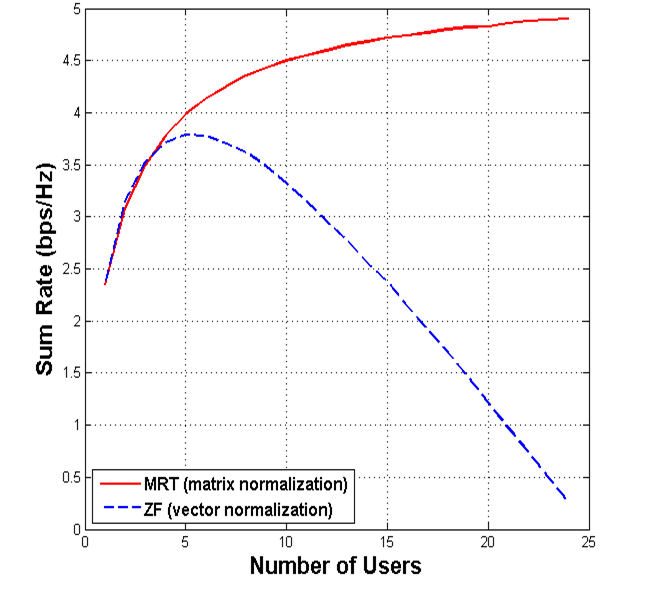}}
    \vertfig[Achievable rate at uplink.]{\label{d}\includegraphics[width=0.5\textwidth]{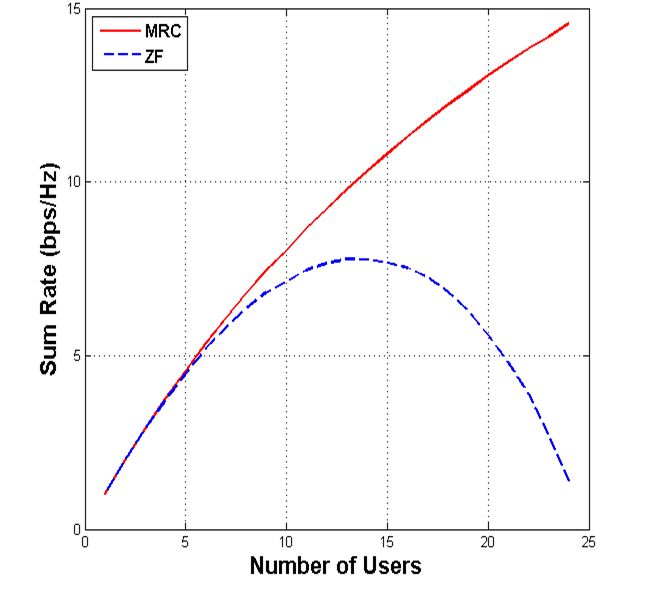}}
    \caption{Achievable rate vs total SNR, where $M$ = 24, $K$ = [1, 24], (a) $P_\text{t} = P_\text{cross,DL} = -7.6$~dB, (b) $P_\text{u} = P_\text{cross,UL} = -13.6$~dB.}
\label{cd}
\end{figure}
\begin{figure}[!t]
 \centering
    \vertfig[Achievable rate at downlink.]{\label{0dB}\includegraphics[width=0.5\textwidth]{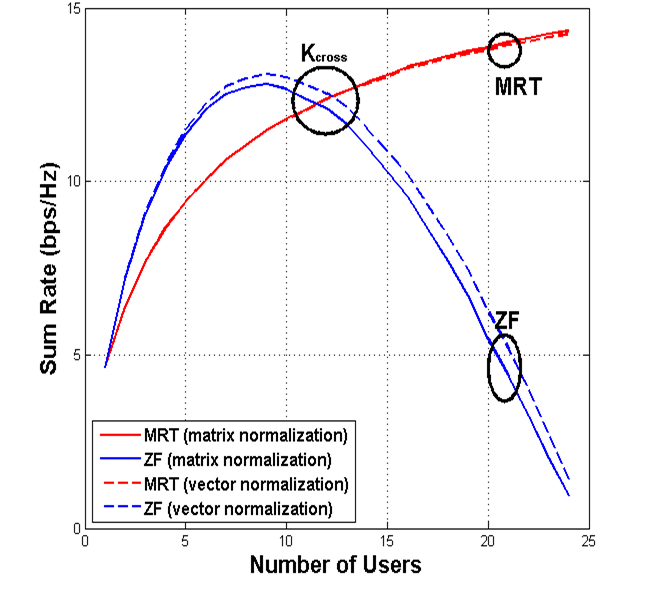}}
    \vertfig[Achievable rate at downlink.]{\label{-5dB}\includegraphics[width=0.5\textwidth]{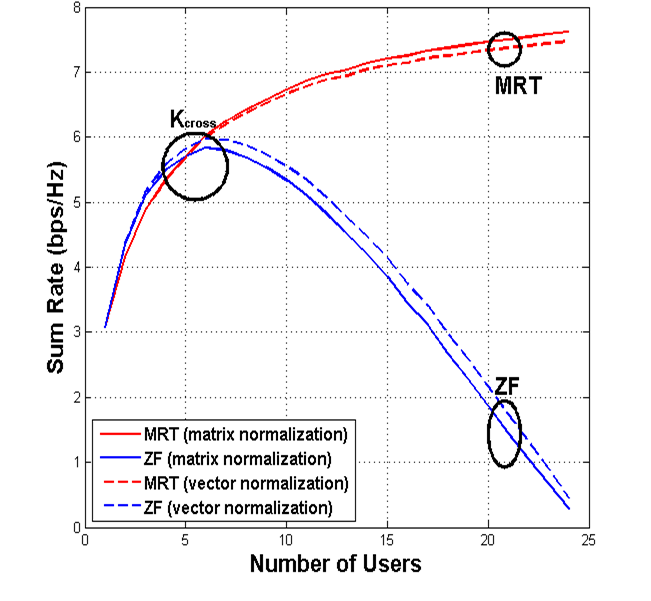}}
    \caption{Achievable rate vs. the number of cell-boundary users, where $M$ = 24, $K$ = [1, 24], and total SNR = (a) $0$~dB, (b) $5$~dB.}
\label{0_5dB}
\end{figure}

\begin{figure}[!t]
 \centering
    \vertfig[Achievable rate of MRT at low SNR.]{\label{MRTvsM}\includegraphics[width=0.5\textwidth]{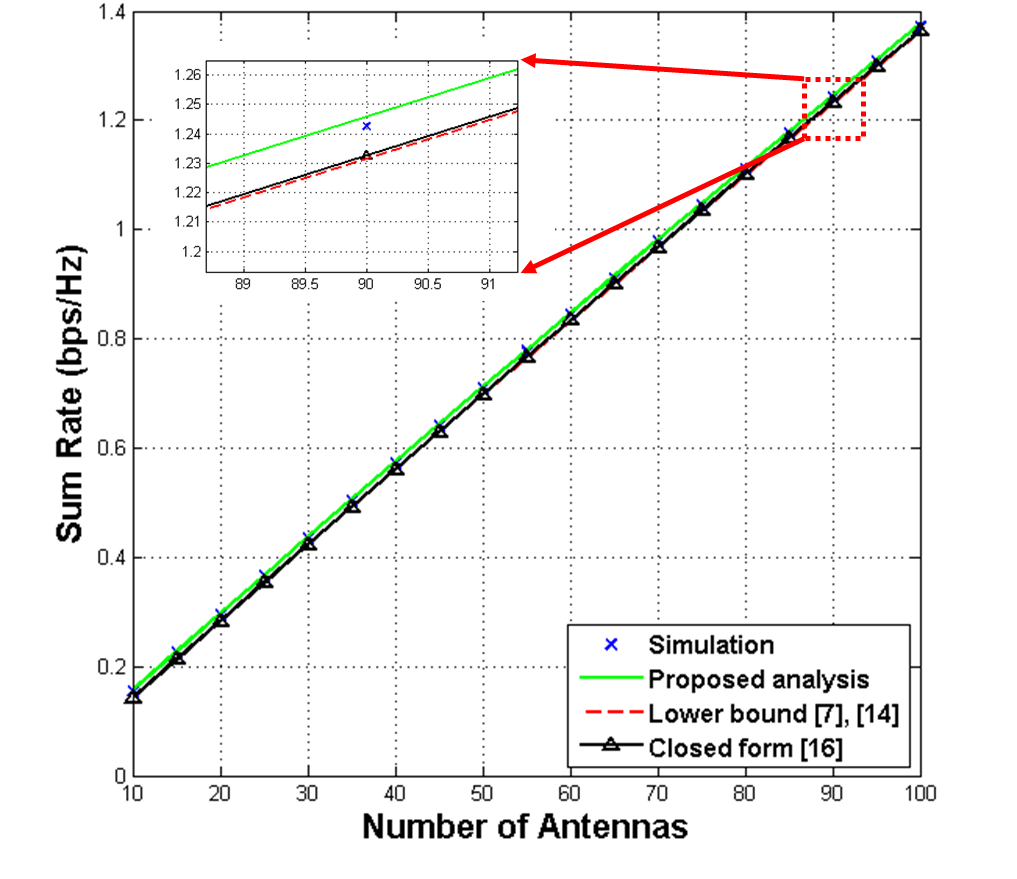}}
    \vertfig[Achievable rate of MRC at low SNR.]{\label{MRCvsM}\includegraphics[width=0.5\textwidth]{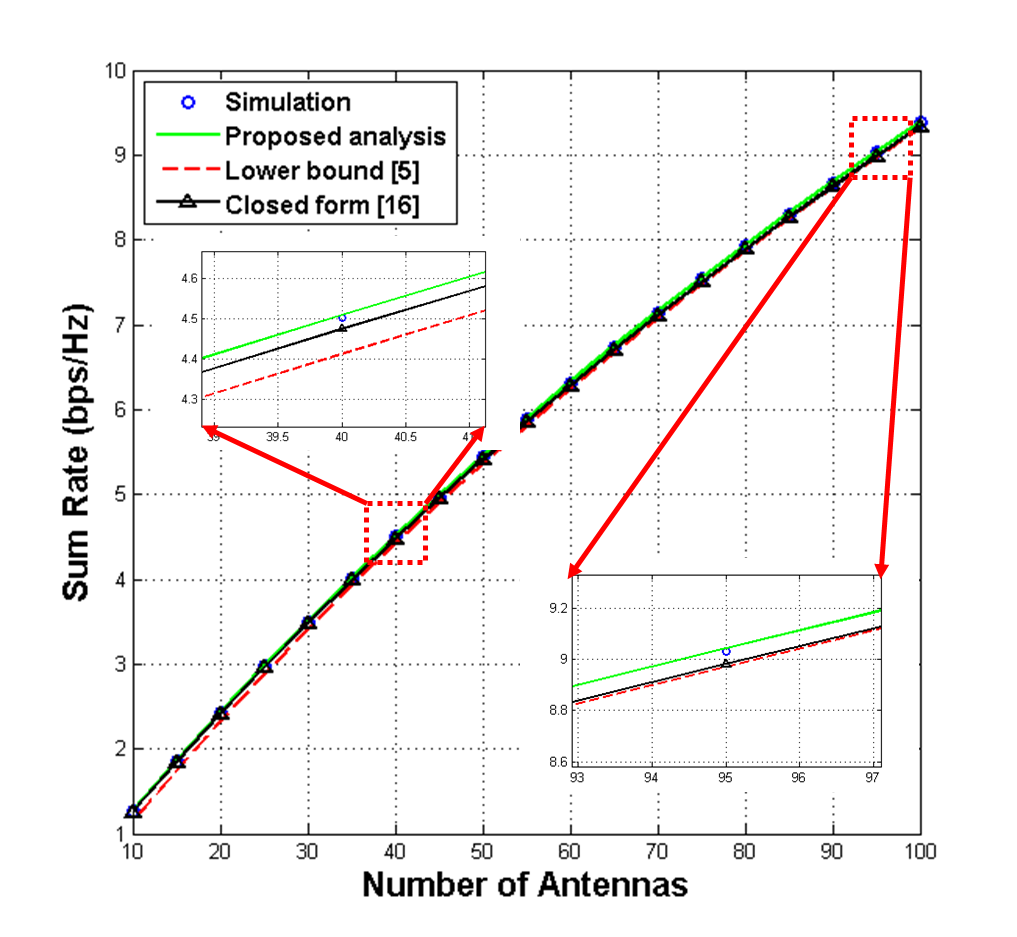}}
    \caption{Achievable rate vs. the number of antennas at the BS, where $K$ = 10, $M$ = [10, 100], and total SNR = -20 dB.}
\label{MRT_MRCvsM}
\end{figure}

For numerical comparisons, we assume that each RU has eight transmit antennas; thus the cloud BS has a total of 24 antennas. \emph{Note that any number of antennas can be used and this constraint is not really related to our system.} This assumption is based on {\color{black}the parameters of 3GPP LTE-advanced}; {\color{black}Release~10} supports eight Node B antennas \cite{Baker_LTE12}. 

Fig. {\color{black}\ref{ZFMRT}(a)} shows the achievable sum rate of ZF for downlink at low SNR. We compare the simulation results with their theoretical upper bound. As mentioned in Section~\ref{Section3}, the ergodic achievable sum rate of ZF with vector normalization {\color{black}approaches its upper bound} at low SNR while that of ZF with matrix normalization approaches its first upper bound. Note that {\color{black}the} \emph{first upper bound} is obtained by Monte Carlo's simulation because $\mathbb{E}\left\{\frac{1}{||F||_F^2}\right\}$ is unknown. Fig.~{\color{black}\ref{ZFMRT}(b)} also describes the achievable sum rate of MRT with $1/M$ ($-13.8$~dB) total SNR. Our achievable sum rate result is almost the same as the numerical results for vector normalization. There is, however, a gap between the simulation results and the proposed achievable sum rate for matrix normalization. This is because the proposed achievable sum rate form is accurate when {\color{black}SNR is lower than $1/M$}. {\color{black}Our analysis is more accurate than the closed forms in \cite{Hoon11,Antonia, Jakob}, which are the same as the lower bound of ZF. In addition, our expression of ergodic sum rate is more accurate than the closed forms in \cite{Hoon11, Jakob} that are given by $K\log_2\left\{1+\frac{P_\text{t}M}{P_\text{t}K+K}\right\}$ and these could be the lower bounds of our expression.} From this comparison, we could confirm (as was also shown in Section~\ref{Section3}) that ZF with vector normalization is better. In contrast, MRT with matrix normalization is better at getting an improved sum rate performance at low SNR.

Figs.~{\color{black}\ref{f3f4}(a)} and {\color{black}\ref{f3f4}(b)} show that the results from (\ref{eq4_3}) and (\ref{eq4_4}) are approximately the same as the ergodic achievable uplink sum rate of MRC where $P_\text{u} \ge M$ and $P_\text{u} \le 1/M$, respectively. Note the large gap between the ergodic achievable uplink sum rate and the lower bound of MRC with finite $M$ shown in \cite{Ngo_TCOM12}. The legend, \emph{Simulation}, indicates the ergodic achievable uplink sum rate of MRC (\ref{eq4_2}) while the proposed analysis in Figs.~{\color{black}\ref{f3f4}(a)} and {\color{black}\ref{f3f4}(b)} indicates the approximation shown in (\ref{eq4_3}) and (\ref{eq4_4}), respectively. Note that the lower bound in \cite{Ngo_TCOM12} is given by $ \mathcal{R}_\text{MRC}^{L}=K\log_2\left\{1+\frac{P_\text{u}(M-1)}{P_\text{u}(K-1)+1}\right\}$ and the closed form in \cite{Jakob} is given by $\mathcal{R}_\text{MRC}=K\log_2\left\{1+\frac{P_\text{u}M}{P_\text{u}K+1}\right\}.$
{\color{black}In addition, the sum rate approaches to $M$ when $M=K$ at high SNR, and the sum rate approaches to $M\log_2\left\{1+\frac{P_\text{u, sum}}{P_\text{u, sum}+1}\right\}$ when $M=K$ at low SNR. This shows that the results of (\ref{MF_largeK}) and (\ref{MF_largeK_sc}) could hold even $M$ is finite.}

In Fig. {\color{black}\ref{ab}(a)}, we illustrate the achievable sum rate of MRT precoding and ZF precoding at downlink. Fig.~{\color{black}\ref{ab}(b)} illustrates the achievable sum rate of MRC and ZF at uplink ($P_\text{th}$ = $6.2$~dB and {\color{black}$-7$~dB are} calculated by \emph{Lemma}s \ref{lemma1} and \ref{lemma2} with $M=24$ and $K=20$ at downlink and uplink, respectively). Note that a cross point of MRT (or MRC) curve and ZF curve is the power threshold. 
{\color{black}It is well known that MRT/MRC performs better than ZF at low SNR. This is the same in the massive MIMO systems, and we can easily find a borderline between MRT/MRC region and ZF region with the proposed simple equation.}
Figs.~{\color{black}\ref{cd}(a)} and {\color{black}\ref{cd}(b)} show that MRT/MRC is always better than ZF at very low SNR regardless of $K$. Used for simulations were $P_\text{t} = P_\text{cross,DL} = -7.6$~dB at downlink and {\color{black}$P_\text{u} = P_\text{cross,UL} = -13.6$~dB at uplink}, and $M = 24$. This result verifies \emph{Lemma}~\ref{lemma3}. We summarize also our conclusions in Tables~\ref{Table1}, \ref{Table2}, and \ref{Table3}.

In Fig. {\color{black}\ref{0_5dB}(a)}, we also compare the achievable sum rates of ZF precoding with MRT precoding when the total transmit SNR is $0$~dB. It shows that ZF with vector normalization is better while MRT with matrix normalization is better for achieving higher sum rates. Fig.~{\color{black}\ref{0_5dB}(b)} illustrates the achievable sum rates of ZF- and MRT-precoding with $-5$~dB transmit SNR. The result is similar to that found in Fig.~{\color{black}\ref{0_5dB}(a)}. As mentioned in Section~\ref{Section5}, in the low SNR regime, using MRT precoding is generally better when the number of active users is larger than $K_{\text{cross, DL}}$. Also, we realize through Figs.{\color{black}~\ref{0_5dB}(a) and \ref{0_5dB}(b)} that as SNR decreases, $K_{\text{cross, DL}}$ shifts to the left. This means that the cloud BS could determine a precoding by $K_{\text{cross, DL}}$ at low SNR.

The performance of MRT/MRC increases as $K$ increases while the performance of ZF decreases as $K$ increases. This is because the ergodic sum rate of ZF at $M=K$ goes to a very small constant at downlink, as shown in (\ref{MK}). Similarly, at uplink, the ergodic sum rate of ZF at $M=K$ could also close to a very small value since its lower bound is zero.

{\color{black}{\color{black}Figs.~\ref{MRT_MRCvsM}(a) and \ref{MRT_MRCvsM}(b)} show that the achievable sum rate of MRT/MRC at low SNR. To predict {\color{black}the} tightness of our analysis for infinite $M$, we compare the simulation result of our analysis with that of the closed form in \cite{Jakob} and the lower bound in {\color{black}\cite{Rusek_SPMag_12} and \cite{Yang}} in the downlink scenario with a large number of antennas at the BS (up to $M=100$). Similarly, we compare the simulation result of our analysis with that of the closed form in \cite{Jakob} and the lower bound in \cite{Ngo_TCOM12} in the uplink scenario. Both numerical results of MRT/MRC show that{\color{black},  when $M$ goes to infinity,} our analyses are quite tighter than {\color{black}the} work given in \cite{Ngo_TCOM12,Rusek_SPMag_12, Yang, Jakob}. Note that {\color{black}the} closed form in \cite{Jakob} is very tight when $M=40$ but {\color{black}as $M$ increases not as tight as} our analysis in the low SNR regime and the uplink scenario. {\color{black}The computation complexity is the same as the lower bounds in \cite{Ngo_TCOM12,Rusek_SPMag_12,Yang} but the accuracy is better than the closed form in \cite{Jakob}. Therefore, we conclude that the proposed ergodic
sum rate is tighter and simpler than the previous work.}}

\begin{table}[!t]
\caption{Precoding normalization techniques in network massive MIMO systems.}
\begin{center}
\begin{tabular}{|c||c|}
\hline
& Precoding normalization technique\\
\hline \hline
$\text{ZF}$ & Vector normalization $\ge$ Matrix normalization \\
\hline
$\text{MRT}$ & Matrix normalization $\ge$ Vector normalization \\
 \hline
\end{tabular}
\end{center}
\label{Table1}
\end{table}
\begin{table}[!t]
\caption{Optimal switching point in network massive MIMO systems.} 
\begin{center}
\begin{tabular}{|c||c|c|}
\hline
& $P_\text{th}$ & $K_{\text{cross}}$ \\
\hline \hline
Downlink & $\frac{K^2}{(K-1)(M-K+1)}$ & $\frac{P_\text{t}(M+1)}{1+P_\text{t}}$\\
\hline
Uplink & $\frac{1}{M-K+1}$ & $M+1-\frac{1}{P_\text{u}}$ \\
 \hline
\end{tabular}
\end{center}
\label{Table2}
\end{table}
\begin{table}[!t]
\caption{Desired technique in network massive MIMO systems. ZF $(\text{if}~K \leq K_{\text{cross}})$ and MRT/MRC $(\text{if}~K \geq K_{\text{cross}})$.}
\begin{center}
\begin{tabular}{|c||c|c|}
\hline
& $K_{\text{cross}}$ & Precoding technique \\
\hline \hline
Cell-center & Large & Zero-forcing \\
\hline
Cell-boundary & Small & MRT/MRC \\
 \hline
\end{tabular}
\end{center}
\label{Table3}
\end{table}

\section{Conclusions}\label{Section7}
In this paper, we proposed massive MIMO systems supporting multiple cell-boundary UEs. For precoding designs, we first derived the achievable sum rate bounds of zero-forcing (ZF) and the approximation of the ergodic achievable sum rate of maximum ratio transmission (MRT) with vector/matrix normalization. Through analytical and numerical results, we confirmed that vector normalization is better for ZF {\color{black}and that} matrix normalization is better for MRT. We also investigated the optimal mode{\color{black}-}switching point as functions of the power threshold and the number of active users in a network. According to the mathematical and numerical results the BS can select a transceiver mode to increase the sum rate for both downlink and uplink scenarios. We {\color{black}anticipate} our analysis {\color{black}providing} insights {\color{black}for related studies, such as those on performance analysis of MIMO, power normalization methods, and the trade-off between ZF and MRT/MRC.} In future work, we will consider {\color{black}more practical scenarios including} limited cooperation among RUs and cooperation delay.


%

\appendix
\subsection{Proof of Lemma 1}\label{Appendix Pre}
Let $h_{k, m}$ be the $m$-th element of $\pmb{h}_k$. Since $h_{k, m}$ is an i.i.d. complex Gaussian random variable with zero mean and unit variance, i.e., $h_{k, m}\sim\mathcal{CN}(0,1)$, $|h_{k, m}|^2$ is a Gamma random variable with unit shape parameter and unit scale parameter, i.e., $|h_{k, m}|^2\sim\Gamma(1,1)$ from the relationship between Rayleigh distribution and Gamma distribution. Therefore, we can say that $|h_{k, m}|^2$ is an exponential random variable with unit parameter ($\lambda =1$), i.e., $|h_{k, m}|^2\sim\text{Exp}(1)$ by the property of Gamma distribution and exponential distribution. Since the $n$-th moment of the exponential random variable is $\frac{n!}{\lambda}$, we can obtain $\mathbb{E}(|h_{k, m}|^4)=2$, $\mathbb{E}(|h_{k, m}|^6)=6$, and $\mathbb{E}(|h_{k, m}|^8)=24$.

\subsubsection{Proof of Lemma 1. 3)}
The expectation of $||\pmb{h}_k||^4$ is given by 
{\color{black}
\begin{align}
\mathbb{E}[||\pmb{h}_k||^4]&=\mathbb{E}\{(|h_{k,1}|^2+|h_{k,2}|^2+\cdots+|h_{k,M}|^2)^2\}\nonumber\\
&=M\mathbb{E}(|h_{k,m}|^4)+\Perms{M}{2}\mathbb{E}(|h_{k,m}|^2|h_{k,i}|^2)\nonumber\\
&=2M+M(M-1)=M^2+M,\nonumber
\end{align}
where $i\neq m$.} The notation $\Perms{n}{r}$ denotes a permutation {\color{black}operator}. The variance of $||\pmb{h}_k||^4$ is also derived by
{\color{black}
\begin{align}
&\text{Var}[||\pmb{h}_k||^4]\nonumber\\
&=M\text{Var}(|h_{k,m}|^4)+\Perms{M}{2}\text{Var}(|h_{k,m}|^2|h_{k,i}|^2)\nonumber\\
&+\Perms{M}{2}\text{Cov}(|h_{k,m}|^2|h_{k,i}|^2, |h_{k,i}|^2|h_{k,m}|^2)\nonumber\\
&+4\Perms{M}{2}\text{Cov}(|h_{k,m}|^4, |h_{k,i}|^2|h_{k,m}|^2)\nonumber\\
&+4\Perms{M}{3}\text{Cov}(|h_{k,m}|^2|h_{k,i}|^2, |h_{k,m}|^2|h_{k,j}|^2)\nonumber\\
&=20M+3M(M-1)+3M(M-1)\nonumber\\
&+16M(M-1)+4M(M-1)(M-2)\nonumber\\
&=4M^3+10M^2+6M.
\label{proof var hk4}
\end{align}}
The variance terms and the covariance terms in (\ref{proof var hk4}) are calculated by
\begin{align}
\begin{split}
&\text{Var}(|h_{k,m}|^4)=20,\\
&\text{Var}(|h_{k,m}|^2|h_{k,i}|^2)=3,\\
&\text{Cov}(|h_{k,m}|^2|h_{k,i}|^2, |h_{k,i}|^2|h_{k,m}|^2)=3,\\
&\text{Cov}(|h_{k,m}|^4, |h_{k,i}|^2|h_{k,m}|^2)=4,\\
&\text{Cov}(|h_{k,m}|^2|h_{k,i}|^2, |h_{k,m}|^2|h_{k,j}|^2)=1 \nonumber
\end{split}
\end{align}
where $i\neq j\neq m$.

\subsubsection{Proof of Lemma 1. 4)}
The expectation of {\color{black}$|\pmb{h}_k^*\pmb{h}_\ell|^2$} is
\begin{align}
\mathbb{E}[|\pmb{h}_k^*\pmb{h}_\ell|^2]&=\mathbb{E}(|h_{k, 1}^*h_{\ell, 1}+h_{k, 2}^*h_{\ell, 2}+\cdots+h_{k, M}^*h_{\ell, M}|^2)\nonumber\\
&=M\mathbb{E}(|h_{k, m}^*h_{\ell, m}|^2)+\Perms{M}{2}\mathbb{E}(h_{k, m}^*h_{\ell, m}h_{k, i}^*h_{\ell, i})\nonumber\\
&=M\nonumber
\end{align}
where $i\neq m$. The variance of {\color{black}$|\pmb{h}_k^*\pmb{h}_\ell|^2$} is also expressed as
{\color{black}
\begin{align}
\text{Var}[|\pmb{h}_k^*\pmb{h}_\ell|^2]&=\text{Var}(|h_{k, 1}^*h_{\ell, 1}+h_{k, 2}^*h_{\ell, 2}+\cdots+h_{k, M}^*h_{\ell, M}|^2)\nonumber\\
&=M\text{Var}(|h_{k, m}^*h_{\ell, m}|^2)\nonumber\\
&+\Perms{M}{2}\text{Var}(h_{k, m}^*h_{\ell, m}h_{k, i}^*h_{\ell, i})\nonumber\\
&=3M+M(M-1)=M^2+2M
\label{proof var hkhl2}
\end{align}}
where $\text{Var}(|h_{k, m}^*h_{\ell, m}|^2)$ is 3, and $\text{Var}(h_{k, m}^*h_{\ell, m}h_{k, i}^*h_{\ell, i})$ is 1 ($i\neq m$). Note that the covariance terms in (\ref{proof var hkhl2}) are all zeros.
\subsection{Proof of Ergodic Sum Rate of MRT}
First, we suppose {\color{black}that $\text{Var}[P_{\text{t}}||\pmb{h}_k||^2]$ and $\text{Var}[\frac{1}{P_{\text{t}}}||\pmb{h}_k||^2]$ converge to zero at low and high SNR, respectively, to evaluate the ergodic achievable sum rate of vector normalization. We can then obtain the following equations:}
\begin{align}
P_\text{t}||\pmb{h}_k||^2\approx P_\text{t}M\label{low lemma}
\end{align}
at low SNR, and
\begin{align}
\frac{1}{P_\text{t}}||\pmb{h}_k||^2\approx\frac{1}{P_\text{t}}M
\label{high lemma}
\end{align}
at high SNR from \emph{Lemmas} \ref{pre1} and (\ref{pre4}).

Next, we assume {\color{black}that $\text{Var}[P_{\text{t}}||\pmb{h}_k||^4]$ and $\text{Var}[|\pmb{h}_k^*\pmb{h}_\ell|^2]$ converge to zero at high SNR to evaluate the ergodic achievable sum rate of matrix normalization. Also, $\text{Var}[\frac{1}{P_\text{t}}||\pmb{H}||_F^2]$ converges to zero at high SNR. Similar to the vector normalization case, we can obtain the following equations:}
\begin{align}
P_\text{t}||\pmb{h}_k||^4\approx P_\text{t}(M^2+M),~~P_\text{t}|\pmb{h}_k^*\pmb{h}_\ell|^2\approx P_\text{t}M\label{low lemma2}
\end{align}
at low SNR, and
\begin{align}
\frac{1}{P_\text{t}}||\pmb{H}||_F^2=\frac{1}{P_\text{t}}\sum_{k=1}^K |\pmb{h}_k|^2\approx\frac{1}{P_\text{t}}KM
\label{high lemma2}
\end{align}
at high SNR.

\subsubsection{Proof of (\ref{MRT vector low SNR})}\label{Appendix B1}
{\color{black}
\begin{align}
\mathcal{R}_{\text{MRT}_{\text{vec}}, \text{ DL}_{\text{L}}}&=\mathbb{E}\left[\sum_{k=1}^{K}\log_2\left\{1+\frac{P_\text{t}\frac{||\pmb{h}_k||^4}{||\sqrt{K}\pmb{h}_k||^2}}{P_\text{t}\sum_{\ell=1,\ell\neq k}^{K}\frac{|\pmb{h}_k^{*}\pmb{h}_\ell|^2}{||\sqrt{K}\pmb{h}_\ell||^2}+1}\right\}\right]\nonumber\\
&=K\mathbb{E}\left[\log_2\left\{1+\frac{P_\text{t}\frac{||\pmb{h}_k||^2}{K}}{P_\text{t}\sum_{\ell=1,\ell\neq k}^{K}\frac{|\pmb{h}_k^{*}\pmb{h}_\ell|^2}{K||\pmb{h}_\ell||^2}+1}\right\}\right]\nonumber\\
&\overset{(a)}{\approx}K\mathbb{E}\left[\log_2\left\{1+\frac{P_\text{t}\frac{M}{K}}{P_\text{t}\sum_{\ell=1,\ell\neq k}^{K}\frac{|\pmb{h}_k^{*}\pmb{h}_\ell|^2}{K||\pmb{h}_\ell||^2}+1}\right\}\right]\nonumber\\
&\overset{(b)}{\approx}K\log_2\left\{1+\frac{P_\text{t}\frac{M}{K}}{P_\text{t}\sum_{\ell=1,\ell\neq k}^{K}\frac{\mathbb{E}(|\pmb{h}_k^{*}\pmb{h}_\ell|^2)}{\mathbb{E}(K||\pmb{h}_\ell||^2)}+1}\right\}\nonumber\\
&\overset{(c)}{=}K\log_2\left\{1+\frac{P_\text{t}M}{P_\text{t}(K-1)+K}\right\}\nonumber
\end{align}
where ($a$) results from (\ref{low lemma}).\footnote{{\color{black}Since $\text{Var}[||\pmb{h}_k||^2](=M)<\text{Var}|\pmb{h}_k^*\pmb{h}_\ell|^2](=M^2+2M)$ from \emph{Lemma}~\ref{pre1}, $\text{Var}[P_{\text{t}}||\pmb{h}_k||^2]$ fastly converges to zero than $\text{Var}[P_{\text{t}}|\pmb{h}_k^*\pmb{h}_\ell|^2]$ at low SNR. We use this property for the rest of the proofs.}} Approximation ($b$) and equality ($c$) can be obtained by \emph{Lemma}s~\ref{pre5} and \ref{pre1}, respectively.} 

\subsubsection{Proof of (\ref{MRT vector high SNR})}\label{Appendix B2}
\begin{align}
\mathcal{R}_{\text{MRT}_{\text{vec}}, \text{ DL}_{\text{H}}}&=\mathbb{E}\left[\sum_{k=1}^{K}\log_2\left\{1+\frac{P_\text{t}\frac{||\pmb{h}_k||^4}{||\sqrt{K}\pmb{h}_k||^2}}{P_\text{t}\sum_{\ell=1,\ell\neq k}^{K}\frac{|\pmb{h}_k^{*}\pmb{h}_\ell|^2}{||\sqrt{K}\pmb{h}_\ell||^2}+1}\right\}\right]\nonumber\\
&=K\mathbb{E}\left[\log_2\left\{1+\frac{P_\text{t}\frac{||\pmb{h}_k||^4}{K||\pmb{h}_k||^2}}{P_\text{t}\sum_{\ell=1,\ell\neq k}^{K}\frac{|\pmb{h}_k^{*}\pmb{h}_\ell|^2}{K||\pmb{h}_\ell||^2}+1}\right\}\right]\nonumber\\
&\overset{(d)}{\approx}K\mathbb{E}\left[\log_2\left\{1+\frac{P_\text{t}\frac{||\pmb{h}_k||^4}{KM}}{P_\text{t}\sum_{\ell=1,\ell\neq k}^{K}\frac{|\pmb{h}_k^{*}\pmb{h}_\ell|^2}{KM}+1}\right\}\right]\nonumber\\
&\overset{(e)}{\approx}K\log_2\left\{1+\frac{P_\text{t}(M+1)}{P_\text{t}(K-1)+K}\right\}\nonumber
\end{align}
where ($d$) results from (\ref{high lemma}) and ($e$) can be obtained by \emph{Lemma}s~\ref{pre1} and \ref{pre5}. 
\subsubsection{Proof of (\ref{MRT matrix low high SNR})}\label{Appendix B3}
in the low SNR regime
\begin{align}
&\mathcal{R}_{\text{MRT}_{\text{mat}}, \text{ DL}_{\text{L}}}\nonumber\\
&=\mathbb{E}\left[\sum_{k=1}^{K}\log_2\left\{1+\frac{P_\text{t}\frac{||\pmb{h}_k||^4}{||\pmb{H}||_F^2}}{P_\text{t}\sum_{\ell=1,\ell\neq k}^{K}\frac{|\pmb{h}_k^{*}\pmb{h}_\ell|^2}{||\pmb{H}||_F^2}+1}\right\}\right]\nonumber\\
&=K\mathbb{E}\left[\log_2\left\{1+\frac{P_\text{t}||\pmb{h}_k||^4}{P_\text{t}\sum_{\ell=1,\ell\neq k}^{K}|\pmb{h}_k^{*}\pmb{h}_\ell|^2+||\pmb{H}||_F^2}\right\}\right]\nonumber\\
&\overset{(f)}{\approx}K\mathbb{E}\left[\log_2\left\{1+\frac{P_\text{t}(M^2+M)}{P_\text{t}\sum_{\ell=1,\ell\neq k}^{K}M+||\pmb{H}||_F^2}\right\}\right]\nonumber\\
&\overset{(g)}{\approx}K\log_2\left\{1+\frac{P_\text{t}(M+1)}{P_\text{t}(K-1)+K}\right\}\nonumber
\end{align}
where ($f$) can be obtained by using (\ref{low lemma2}) directly. ($g$) can also be derived by \emph{Lemma}~\ref{pre5} and $\mathbb{E}(||\pmb{H}||_F^2)=MK$.
\subsubsection{Proof of (\ref{MRT matrix low high SNR})}\label{Appendix B4}
in the high SNR regime
\begin{align}
&\mathcal{R}_{\text{MRT}_{\text{mat}}, \text{ DL}_{\text{H}}}\nonumber\\
&=\mathbb{E}\left[\sum_{k=1}^{K}\log_2\left\{1+\frac{P_\text{t}\frac{||\pmb{h}_k||^4}{||\pmb{H}||_F^2}}{P_\text{t}\sum_{\ell=1,\ell\neq k}^{K}\frac{|\pmb{h}_k^{*}\pmb{h}_\ell|^2}{||\pmb{H}||_F^2}+1}\right\}\right]\nonumber\\
&=K\mathbb{E}\left[\log_2\left\{1+\frac{||\pmb{h}_k||^4}{\sum_{\ell=1,\ell\neq k}^{K}|\pmb{h}_k^{*}\pmb{h}_\ell|^2+\frac{1}{P_\text{t}}||\pmb{H}||_F^2}\right\}\right]\nonumber\\
&\overset{(h)}{\approx}K\mathbb{E}\left[\log_2\left\{1+\frac{||\pmb{h}_k||^4}{\sum_{\ell=1,\ell\neq k}^{K}|\pmb{h}_k^{*}\pmb{h}_\ell|^2+\frac{1}{P_\text{t}}MK}\right\}\right]\nonumber\\
&\overset{(i)}{\approx}K\log_2\left\{1+\frac{P_\text{t}(M+1)}{P_\text{t}(K-1)+K}\right\}\nonumber
\end{align}
where ($h$) results from (\ref{high lemma2}) while ($i$) can be obtained by \emph{Lemma}s \ref{pre1} and \ref{pre5} as well.

\subsection{Proof of Ergodic Sum Rate of MRC}\label{Appendix C}
\subsubsection{Proof of (\ref{eq4_3})}
\begin{align}
&\mathcal{R}_{\text{MRC, UL}_{\text{H}}}\nonumber\\
&=\mathbb{E}\left[K\log_2\left\{1+\frac{P_\text{u}||\pmb{h}_k||^4}{P_\text{u}\sum_{\ell=1,\ell\neq k}^{K}|\pmb{h}_k^{*}\pmb{h}_\ell|^2+||\pmb{h}_k||^2}\right\}\right]\nonumber\\
&=\mathbb{E}\left[K\log_2\left\{1+\frac{||\pmb{h}_k||^4}{\sum_{\ell=1,\ell\neq k}^{K}|\pmb{h}_k^{*}\pmb{h}_\ell|^2+\frac{1}{P_\text{u}}||\pmb{h}_k||^2}\right\}\right]\nonumber\\
&\approx \mathbb{E}\left[K\log_2\left\{1+\frac{||\pmb{h}_k||^4}{\sum_{\ell=1,\ell\neq k}^{K}|\pmb{h}_k^{*}\pmb{h}_\ell|^2+\frac{1}{P_\text{u}}M}\right\}\right]\nonumber\\
&\approx K\log_2\left\{1+\frac{P_\text{u}(M+1)}{P_\text{u}(K-1)+1}\right\}
\label{eq4_3 proof}.
\end{align}
using (\ref{low lemma}), (\ref{high lemma}), \emph{Lemma}s \ref{pre1} and \ref{pre5}.

\subsubsection{Proof of (\ref{eq4_4})}
\begin{align}
\begin{split}
&\mathcal{R}_{\text{MRC, UL}_{\text{L}}}\nonumber\\
&=\mathbb{E}\left[K\log_2\left\{1+\frac{P_\text{u}||\pmb{h}_k||^4}{P_\text{u}\sum_{\ell=1,\ell\neq k}^{K}|\pmb{h}_k^{*}\pmb{h}_\ell|^2+||\pmb{h}_k||^2}\right\}\right]\nonumber\\
&=\mathbb{E}\left[K\log_2\left\{1+\frac{P_\text{u}||\pmb{h}_k||^2}{P_\text{u}\sum_{\ell=1,\ell\neq k}^{K}\frac{|\pmb{h}_k^{*}\pmb{h}_\ell|^2}{||\pmb{h}_k||^2}+1}\right\}\right]\nonumber\\
&\approx\mathbb{E}\left[K\log_2\left\{1+\frac{P_\text{u}M}{P_\text{u}\sum_{\ell=1,\ell\neq k}^{K}\frac{|\pmb{h}_k^{*}\pmb{h}_\ell|^2}{||\pmb{h}_k||^2}+1}\right\}\right]\nonumber\\
&\approx K\log_2\left\{1+\frac{P_\text{u}M}{P_\text{u}(K-1)+1}\right\}\nonumber
\end{split}
\end{align}
using the same methods as in (\ref{eq4_3 proof}).


\ifCLASSOPTIONcaptionsoff
  \newpage
\fi



%
%

\bibliographystyle{IEEEtran}
\bibliography{MassiveMIMO_Chae_submitted}


%





\begin{IEEEbiography}{Yeon-Geun Lim}
(S'12) received his B.S. degree in Information and Communications Engineering from Sungkyunkwan University, Korea in 2011. He is now with the School of Integrated Technology, Yonsei University, Korea and is working toward the Ph.D. degree.

His research interest includes massive MIMO and interference management techniques for smart small cell networks. \end{IEEEbiography}

\begin{IEEEbiography}{Chan-Byoung Chae}
(S'06 - M'09 - SM'12) is an Assistant Professor in the School of Integrated Technology, College of Engineering, Yonsei University, Korea. He was a Member of Technical Staff (Research Scientist) at Bell Laboratories, Alcatel-Lucent, Murray Hill, NJ, USA from 2009 to 2011. Before joining Bell Laboratories, he was with the School of Engineering and Applied Sciences at Harvard University, Cambridge, MA, USA as a Post-Doctoral Research Fellow. He received the Ph.D. degree in Electrical and Computer Engineering from The University of Texas (UT), Austin, TX, USA in 2008, where he was a member of the Wireless Networking and Communications Group (WNCG). 

Prior to joining UT, he was a Research Engineer at the Telecommunications R\&D Center, Samsung Electronics, Suwon, Korea, from 2001 to 2005.  While having worked at Samsung, he participated in the IEEE 802.16e standardization, where he made several contributions and filed a number of related patents from 2004 to 2005. His current research interests include capacity analysis and interference management in energy-efficient wireless mobile networks and nano (molecular) communications. He has served/serves as an Editor for the \textsc{IEEE Trans. on Wireless Comm.}, \textsc{IEEE Trans. on Smart Grid} and \textsc{IEEE/KICS Jour. Comm. Nets}. He is a Guest Editor for the \textsc{IEEE Jour. Sel. Areas in Comm.} (special issue on molecular, biological, and multi-scale comm.). He is an IEEE Senior Member.

Dr. Chae was the recipient/co-recipient of the IEEE INFOCOM Best Demo Award (2015), the IEIE/IEEE Joint Award for Young IT Engineer of the Year (2014), the Haedong Young Scholar Award (2013), the IEEE Signal Processing Magazine Best Paper Award (2013), the IEEE ComSoc AP Outstanding Young Researcher Award (2012), the IEEE VTS Dan. E. Noble Fellowship Award (2008), the Gold Prize (1st) in the 14th/19th Humantech Paper Contests, and the KSEA-KUSCO scholarship (2007). He also received the Korea Government Fellowship (KOSEF) during his Ph.D. studies.
\end{IEEEbiography}

\begin{IEEEbiography}{Guiseppe Caire}
(S'92 - M'94 - SM'03 - F'05) was born in Torino, Italy, in 1965. He received the B.Sc. in Electrical Engineering from Politecnico di Torino (Italy), in 1990, the M.Sc. in Electrical Engineering from Princeton University in 1992 and the Ph.D. from Politecnico di Torino in 1994. He was a recipient of the AEI G.Someda Scholarship in 1991, has been with the European Space Agency (ESTEC, Noordwijk, The Netherlands) from May 1994 to February 1995, was a recipient of the COTRAO Scholarship in 1996 and of a CNR Scholarship
in 1997. He has been visiting Princeton University in Summer 1997 and Sydney University in Summer 2000. He has been Assistant Professor in Telecommunications at the Politecnico di Torino, Associate Professor at the University of Parma, Italy, Professor with the Department of Mobile Communications at the Eurecom Institute, Sophia-Antipolis, France, and a professor of Electrical Engineering with the Viterbi School of Engineering, University of Southern California, Los Angeles, CA. He is now a professor with the School of Electrical Engineering and Computer Science, Technical University of Berlin, Germany. He served as Associate Editor for the IEEE TRANSACTIONS ON COMMUNICATIONS in 1998-2001 and as Associate Editor for the IEEE TRANSACTIONS ON INFORMATION THEORY in 2001-2003. He received the Jack Neubauer Best System Paper Award from the IEEE Vehicular Technology Society in 2003, and the IEEE Communications Society \& Information Theory Society Joint Paper Award in 2004 and in 2011. Giuseppe Caire is Fellow of IEEE since 2005. He has served in the Board of Governors of the IEEE Information Theory Society from 2004 to 2007, and as President of the IEEE Information Theory Society in 2011. His main research interests are in the field of communications theory, information theory, channel and source coding with particular focus on wireless communications.
\end{IEEEbiography}

\end{document}